\documentclass[11pt,a4paper]{article}

\pdfoutput=1 % if your are submitting a pdflatex (i.e. if you have
\usepackage{jcappub} % for details on the use of the package, please

\usepackage{bm}
\usepackage{epsfig}
\usepackage{subfigure}

\usepackage{graphicx}% Include figure files

\usepackage{float}
\usepackage{graphicx}
\usepackage{amsmath,amssymb}
\usepackage{textcomp}
\usepackage{gensymb}
\usepackage[utf8]{inputenc}
\usepackage[T1]{fontenc}
\usepackage{setspace}
\usepackage{hyperref}

\def\beq{\begin{equation}}
\def\enq{\end{equation}}
\def\bea{\begin{eqnarray}}
\def\ena{\end{eqnarray}}

\begin{document}

\title{Constraining high-energy neutrino emission from choked jets in stripped-envelope supernovae}

\author{Nicholas Senno,}

\author{Kohta Murase,}

\author{Peter M\'esz\'aros}

\affiliation{Department of Physics; Department of Astronomy \& Astrophysics; Center for Particle and Gravitational Astrophysics, The Pennsylvania State University, University Park, Pennsylvania, 16802, USA}

\abstract{
There are indications that $\gamma$-ray dark objects such as supernovae (SNe) with choked jets, and the cores of active galactic nuclei may contribute to the diffuse flux of astrophysical neutrinos measured by the IceCube observatory. 
In particular, stripped-envelope SNe have received much attention since they are capable of producing relativistic jets and could explain the diversity in observations of collapsar explosions (e.g., gamma-ray bursts (GRBs), low-luminosity GRBs, and Type Ibc SNe).
We use an unbinned maximum likelihood method to search for spatial and temporal coincidences between Type Ibc core-collapse SNe, which may harbor a choked jet, and muon neutrinos from a sample of IceCube up-going track-like events measured from May 2011-May 2012.
In this stacking analysis, we find no significant deviation from a background-only hypothesis using one year of data, and are able to place upper limits on the total amount of isotropic equivalent energy that choked jet core-collapse SNe deposit in cosmic rays ${\mathcal E}_{\rm cr}$ and the fraction of core-collapse SNe which have a jet pointed towards Earth $f_{\rm jet}$. 
This analysis can be extended with yet to be made public IceCube data, and the increased amount of optically detected core-collapse SNe discovered by wide field-of-view surveys such as the Palomar Transient Factory and All-Sky Automated Survey for Supernovae. The choked jet SNe/high-energy cosmic neutrino connection can be more tightly constrained in the near future.
}

\maketitle
\date{\today}

%\pacs{none \vspace{-0.3cm}}
% 95.85.Ry Neutrino, muon, pion, and other elementary particles; cosmic rays

%\maketitle

%%%%%%%%%%%%%%%%%%%%%%%%%%%%%%%%%%%%%%%%%%%%%%%%%%
%%%%%%%%%%%%%%%%%%%%%%%%%%%%%%%%%%%%%%%%%%%%%%%%%%

\section{Introduction}
\doublespacing
The IceCube Antarctic neutrino observatory has been observing high-energy (HE) neutrinos with energies of $E_\nu \gtrsim 10\,{\rm TeV}$ for almost half a decade \cite{2013PhRvL.111b1103A,Aartsen:2013jdh,2014PhRvL.113j1101A,2015PhRvD..91b2001A,Aartsen:2015de,Aartsen:2016xlq}. They observe an apparently isotropic diffuse neutrino flux that is equally distributed between the neutrino flavors (i.e., $\nu_e:\nu_\mu:\nu_\tau \approx1:1:1$) \cite{2015ApJ...809...98A}. 
The sources of these HE neutrinos are still unknown, despite observations of $\sim 100$ HE contained events, and tens of thousands of through-going track events to date. Gamma-ray bright sources such as gamma-ray bursts (GRBs) \cite{1997PhRvL..78.2292W,Murase:2005hy} (see also \cite{Petropoulou:2014lja,Bustamante:2015fb,Bustamante:2016wpu} for the latest papers after the discovery of IceCube neutrinos) and blazars \cite{1995APh.....3..295M,Atoyan:2001ey,Murase:2014foa,Dermer:2014vaa,Tavecchio:2014xha,Padovani:2015mba} are now disfavored as the main origins of IceCube's neutrinos \cite{Aartsen:2014aqy,Murase:2016gly,Aartsen:2016lir}.

During the last few years, $\gamma$-ray dark sources have been investigated as potential sources of the HE diffuse flux (see Ref.~\cite{Murase:2016ck} and references therein). Such sources may be able to accelerate HE cosmic-rays (CRs) which interact with their surrounding environment to produce $\gamma$-rays and neutrinos via $pp$ or $p\gamma$ interactions. While the $\gamma$-rays are attenuated by a dense photon field (via $\gamma\gamma$ interactions), the HE neutrinos escape. 
Studies of $\gamma$-ray dark sources attempt to determine their potential maximum contribution to the diffuse neutrino flux, while still respecting multi-messenger observations such as the diffuse extragalactic $\gamma$-ray background \cite{Murase:2013ffa,Kimura:2014jba,Senno:2015tsn,2016PhRvD..93h3005W,2016PhRvD..93e3010T,Senno:2016bso,2016arXiv161200011D}. By definition, the amount of non-thermal energy in CRs is difficult to directly determine for $\gamma$-ray dark transients, making it harder to constrain the suggested theoretical models (although rare sources such as jetted tidal disruption events (TDEs) are already disfavored by the absence of clustering \cite{Senno:2016bso} and diffuse emission \cite{Senno:2016bso,2016arXiv161200011D,Lunardini:2016wh}). 

It has been known since the combined observation of SN 1998bw/GRB 980425 \cite{1998Natur.395..670G,1999ApJ...516..788W} that so-called long gamma-ray bursts (GRBs) share a common progenitor with core-collapse supernovae (ccSNe).  Prompt GRB emission is observed once a relativistic jet escapes from the stellar envelope of its progenitor during core-collapse. It is then natural to suppose that some jets do not escape the stellar envelope~\cite{2011arXiv1112.5949B,Mizuta:2013he}, instead resulting in a trans-relativistic explosion or stripped-envelope SNe \cite{2013ApJ...778...18M}. While no events have been definitively identified as an example of a ``choked jet'' SN (cjSNe), it is believed that unusually energetic explosions or those with trans-relativistic ejecta such as a Type Ic broad-line SNe or hypernovae may be the result of such a scenario \cite{1999ApJ...516..788W,Nomoto:2001uv,2016ApJ...832..108M}. For recent discussions on the choked jets in core-collpase SNe see \cite{2017MNRAS.472..616S,2017MNRAS.472.1770B}.
Low-luminosity GRBs or trans-relativistic SNe have been suggested as HE neutrino emitters~\cite{Murase:2006mm,Gupta:2006jm,Kashiyama:2012zn}. 

Conventional (i.e., $\gamma$-ray bright) GRBs were thought to be prime candidates to produce HE neutrinos \cite{1997PhRvL..78.2292W}, but they can only contribute to $\lesssim1$\% of IceCube's neutrinos \cite{Abbasi:2011hk,2012Natur.484..351A,Aartsen:2014aqy,Aartsen:2015jp,2016ApJ...824..115A,2017arXiv170206868A}. While it stands to reason that their choked-jet brethren may deposit a significant amount of energy into CRs and neutrinos \cite{2001PhRvL..87q1102M,Razzaque:2003eq,Razzaque:2004ix,2005MPLA...20.2351R,Ando:2005xi,Iocco:2007td}, the non-detection of precursor neutrinos from high-luminosity GRBs already indicate that HE neutrino production should be suppressed inside a star for powerful GRBs.  However, Ref.~\cite{Murase:2013ffa} showed that this is consistent with the theoretical expectation that CR acceleration is inefficient at radiation-mediated shocks, and proposed that low-power choked jets may give a significant contribution to the observed neutrino flux without contradicting the non-detection of GRB neutrinos. Alternatively, Ref.~\cite{Senno:2015tsn} considered HE neutrino production in choked jets embedded in the circumstellar material or extended envelope.  
It has also been suggested that choked jets may account for the observed high-energy neutrinos~\cite{Murase:2013ffa}, especially in light of the medium-energy excess~\cite{Murase:2013ffa,Nakar:2015tma,Senno:2015tsn,2016PhRvD..93e3010T}. 

A search for HE neutrinos coincident with SN 2008D -- assuming it contained a choked jet -- has been performed \cite{2011A&A...527A..28I}. In this current work, we investigate a population of $\gamma$-ray dark transients -- stripped-envelope SNe -- as the potential sources of HE neutrinos. We use observed SNe events to constrain the fraction of the population that can produce neutrinos, and the amount of energy each event can deposit into CRs (${\mathcal E}_{\rm cr}$). The background only hypothesis cannot be ruled out using only the present neutrino data (i.e., there is not a significant association between neutrinos and Type Ibc SNe). We are therefore able to place upper limits on ${\mathcal E}_{\rm cr}$ and the fraction of SNe that could have a choked jet pointed towards Earth ($f_{\rm jet}$). Note that the latter parameter cannot discriminate between SNe that do not have jets, and those which have jets that are not pointing towards us because of beaming considerations. 

The IceCube collaboration is expected to release the remainder of their upgoing track-like events (roughly 7 years in total), which will improve the limits of this analysis. Furthermore, our procedure can be applied to test neutrino coincidences with any $\gamma$-ray dark source (e.g., low-luminosity GRBs, choked-jet TDEs). With the introduction of wide field-of-view optical surveys such as the Palomar Transient Factory \cite{2009PASP..121.1334R} and All-Sky Automated Survey for Supernovae (ASAS-SN)\footnote{\url{http://www.astronomy.ohio-state.edu/~assassin}}, new rich datasets of $\gamma$-ray dark transient events will be made available to test as potential neutrino sources. Furthermore, cjSNe are important targets for the Astrophysical Messenger Observatory Network (AMON)~\cite{Smith:2012eu}. Because current optical telescopes view the entire sky every few days, it will soon be possible to understand the initial period of a SN explosion; and, improving this analysis by determining the exact time at which a choked jet would occur. 

In the following, we take the comosmological parameters to be $H_0 = 68\,{\rm km\,s^{-1}\,Mpc^{-1}}$ with $\Omega_\Lambda = 0.69$ and $\Omega_m = 0.31$.

\section{Data} 

We consider one-year of IceCube data that was taken between May 2011 and May 2012 using the 86 string detector array \cite{Aartsen:2015de}. This sample contains $\sim 70,000$  upgoing, track-like events. To eliminate contamination from atmospheric muons, we consider only upgoing track events. Each neutrino event contains information on its energy proxy -- which is related to the total amount of photoelectrons observed in the detector \cite{2010RScI...81h1101H} -- the day on which it was detected, its arrival direction, and corresponding angular uncertainty. Track-like events typically have median angular uncertainties of $\lesssim 2\degree$, although they depend on their energies. In particular, the low-energy events are affected by large kinematic angles. There are also misreconstructed events. We note that $< 100$ events have large angular uncertainties of $\gtrsim 50\degree$. 

There are 29 Type Ibc SNe \footnote{Note that an additional SN -- SN2012bz \cite{2016ApJ...832..108M} -- is omitted from this analysis because it does not have a measured date on which its optical flux reaches a maximum, and therefore does not meet the requirements of this analysis.} that were detected in the Northern hemisphere during the neutrino data collection period. From the Open SNe catalog (See \cite{Guillochon:2016ci} and references within), we determine the date of their maximum optical brightness, their location in the sky, and redshift (the latter for all but one event). Table \ref{table:sne_data} lists the SNe observations used in our analysis. Note that available catalogues are quite incomplete at present. Also, with the inclusion of new optical transient ``factories'' such as the Palomar Transient Factory, and ASAS-SN the number of Type Ibc SNe detected each year following 2012 is significantly larger. 

\begin{table}
\centering
\begin{tabular}{| l | c | c | c | c | c |} 
\hline
Name & Max Mag (MJD) & RA (rad) & Dec (rad) & $D_L\,({\rm Mpc})$ & Type\\ \hline
SN2011ep & 55750.5 & 4.47 & 0.57 & 1490 & Ic \\ \hline
PTF11ixk & 55765.5 & 3.50 & 0.55 & 95 & Ic \\ \hline 
PTF11izq & 55767.5 & 3.61 & 0.70 & 289 & Ib \\ \hline 
PTF11ilr & 55771.5 & 6.05 & 0.27 & - & Ib \\ \hline 
SN2011ee & 55773.5 & 6.14 & 0.15 & 137 & Ic \\ \hline 
PTF11kaa & 55775.5 & 4.57 & 0.82 & 184 & Ib \\ \hline 
SN2011gd & 55790.5 & 4.34 & 0.38 & 44 & Ib \\ \hline 
PTF11klg & 55810.5 & 5.79 & 0.11 & 120 & Ic \\ \hline 
PFT11kmb & 55820.5 & 5.86 & 0.63 & 77 & Ib-Ca \\ \hline 
SN2011fl & 55829.5 & 0.21 & 0.49 & 71 & Ib \\ \hline 
SN2011ft & 55829.5 & 4.68  & 0.51 & 78 & Ib \\ \hline 
SN2011gh & 55829.5 & 1.97 & 0.45 & 85 & Ib \\ \hline 
PFT11qcj & 55866.5 & 3.46 & 0.83 & 127 & Ib \\ \hline 
SN2011fz & 55888.5 & 6.01 & 0.04 & 73 & Ib \\ \hline 
LSQ11jw & 55909.5 & 0.54. & 0.01 & 90 & Ic \\ \hline 
SN2011jm & 55918.5 & 3.38 & 0.05 & 14 & Ic \\ \hline 
SN2011it & 55919.5 & 5.77 & 0.55 & 72 & Ic \\ \hline 
SN2011kf & 55925.5 & 3.83 & 0.29 & 1280 & Ic \\ \hline 
SN2012C & 55939.5 & 2.52 & 0.57 & 65 & Ic \\ \hline 
SN2012F & 55930.5 & 0.15 &  0.07 & 137 & Ib \\ \hline 
SN2011kg & 55937.5 & 0.44 & 0.52 & 976 & SLSN-I \\ \hline 
SN2012il & 55941.5 & 2.56 & 0.35 & 878& SLSN-I \\ \hline 
SN2012aa & 55954.5 & 3.89 & -0.04 & 376 & Ic \\ \hline 
SN2012ap & 55975.5 & 1.31 & -0.05 & 55 & Ic BL \\ \hline 
PTF12bwq & 56007.5 & 3.61 & 0.44 & 184 & Ib \\ \hline 
PS1-12sk & 56013.5 & 2.29 & 0.75 & 251& Ibn \\ \hline 
LSQ12bph & 56017.5 & 4.13 & 0.4 & 207 & Ic \\ \hline 
SN2012bw & 56039.5 & 4.25 & 0.57 & 141 & Ic \\ \hline 
PTF12cde & 56068.5 &3.66 & 0.63 & 56 & Ib/c \\ \hline 
\end{tabular}
\caption{Observations of the 29 Type Ibc SNe which were detected in the Northern Hemisphere between May 2011 and May 2012.}
\label{table:sne_data}
\end{table}

\section{Analysis} 
\subsection{Signal and Background PDFs}
We use the unbinned maximum log-likelihood method first developed for neutrino astronomy by Braun et al. \cite{2008APh....29..299B} to search for associations of neutrino arrival directions with source positions. This technique was later adapted to analyze the correlation between neutrino events and the prompt emission of GRBs \cite{Aartsen:2015jp}. The same analysis is appropriate for determining the correlation of SNe explosions with neutrino events, although the likelihoods of the neutrino arrival time, direction, and deposited energy must be adjusted for these new potential sources. The method we present can be easily extended to include additional years of IceCube data, or modified for other gamma-ray dark sources, such as choked-jet TDEs. 

The neutrinos in our model are assumed to be produced by a jet that is launched during the core-collapse of the SN progenitor, but is subsequently choked off by the stellar or circumstellar envelope. 
For typical GRB jet luminosities, this process occurs within $\sim 10-100$ s after the initial stellar explosion \cite{2011arXiv1112.5949B,Mizuta:2013he}. However, almost all SN are not detected until days after the initial explosion, and it is hard to precisely reconstruct the stellar explosion time. By studying GRBs that are associated with SNe, it was found that the prompt GRB emission was detected $\sim 13\pm2.3$ days before the SN reached its maximum optical brightness (see \cite{2017AdAst2017E...5C} and references within). 

We therefore use the maximum SN brightness time as a proxy for the stellar explosion time, and assume that for any given Type Ibc SNe, the difference in time between the two is given by a Poisson distribution with $\lambda_T=13$ days. Note that since the arrival times for the neutrinos are coarsely binned by the day they were detected, a discrete probability distribution is appropriate as an approximation here. The signal probability mass function (i.e., the analog of a probability density function but for discrete values) for the arrival time of neutrinos is therefore

\beq
\label{eq:sigt}
S_T (T_{\rm arr,\nu},T_{\rm max,sn})= e^{-\lambda_T}\frac{\lambda_T^{T_{\rm max,sn}-T_{\rm arr,\nu}}}{(T_{\rm max,sn}-T_{\rm arr,\nu})!},
\enq
where $T_{\rm max,sn}$ and $T_{\rm arr,\nu}$ are integers rounded to the nearest Modified Julian Day. 
 
The background neutrino events for this analysis are primarily neutrinos produced by CR interactions in the Earth's atmosphere, and are assumed to occur at a constant rate. Therefore, the background probability density function (PDF) for the arrival time of neutrinos (denoted as $B_{T}$) is constant during the observation window. This window is taken to be the central 99\% confidence interval for Eq. \ref{eq:sigt}, $T_{\rm max,sn} - 19\,{\rm days} \leq T \leq T_{\rm max,sn} - 4 \,{\rm days}$. 

Unlike Ref.~\cite{2008APh....29..299B}, we assume that the signal PDF for neutrino arrival directions is the von Mises-Fisher (aka the Kent \cite{kent1982fisher}) distribution \cite{2016ApJ...816...75A,2017arXiv170206868A}

\beq
\label{eq:sigdir}
S_{\rm dir}\left((\alpha,\delta)_\nu,(\alpha,\delta)_{\rm sn}\right) = \frac{\kappa}{4\pi\sinh\kappa}e^{\kappa\mu}, 
\enq
where $\kappa = 1/\sigma_\nu^2$ is related to the angular error of the neutrino arrival direction, since it can be assumed that the uncertainty of the SN position is negligible compared to $\sim 1\degree$ . The cosine of the angular separation between the neutrino and SN position is $\mu = \cos\Delta\psi$. Note that for $\Delta\psi \ll 1$ and $\kappa \gg 1$ Eq. \ref{eq:sigdir} reduces to a 2D Gaussian PDF $S_{\rm dir} = \frac{1}{2\pi\sigma_\nu^2}e^{-\Delta\psi^2/2\sigma_\nu^2}$ (compare with Eq. 9 in \cite{2008APh....29..299B} and Eq. 3 in \cite{Aartsen:2015jp}). 

\begin{figure}[h]
\centering
\includegraphics[width=0.8\textwidth]{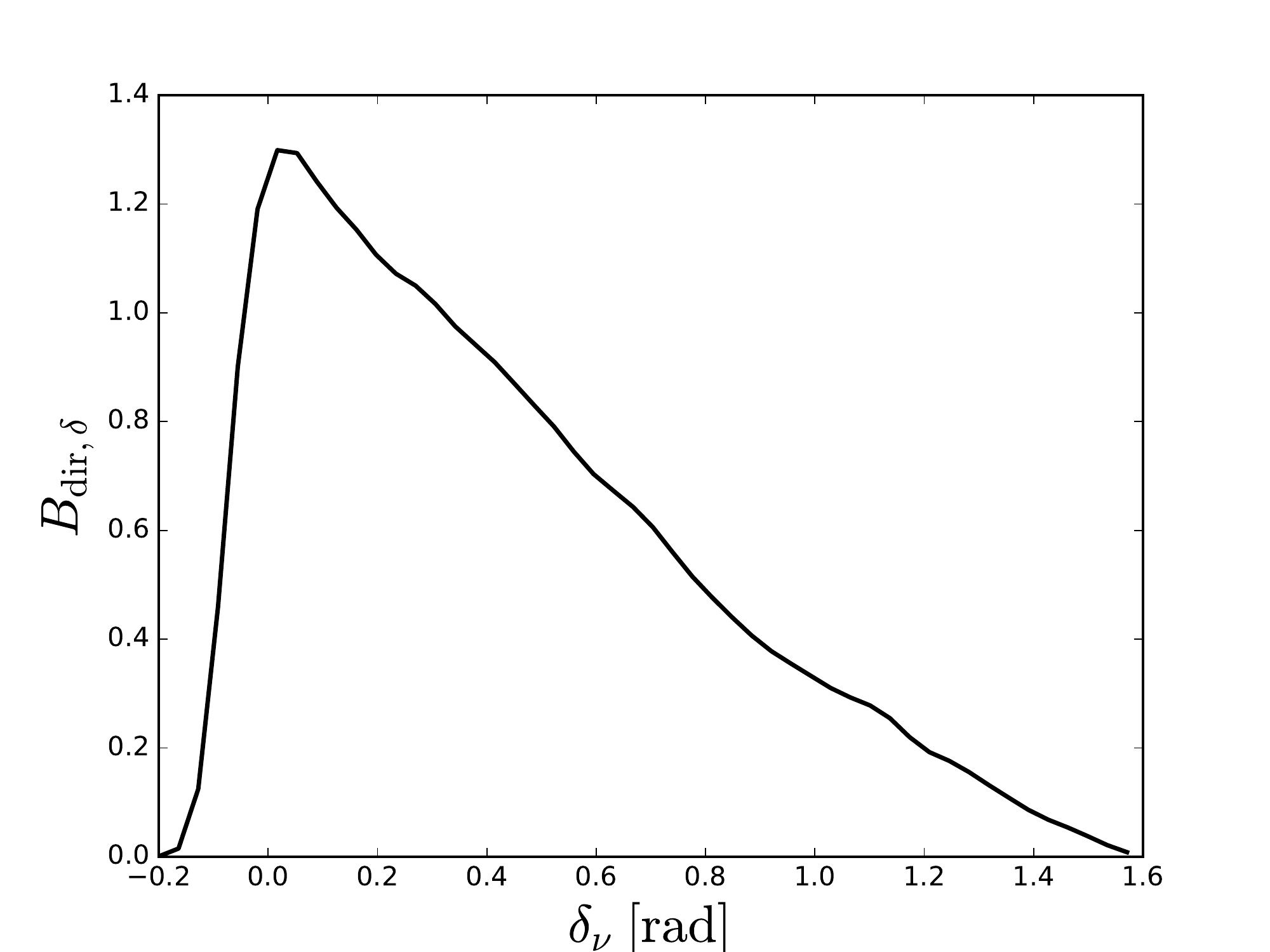}
\caption{Background PDF for neutrino arrival declination. Constructed using neutrino data from events not within the acceptance window (i.e., outside of the 99\% central (lower) confidence level of Eq. \ref{eq:sigt} (\ref{eq:sigdir}))
}
\label{fig:bkgdir} 
\vspace{-1.\baselineskip}
\end{figure}

The background PDF for the arrival direction of neutrinos (denoted as $B_{\rm dir}$) is the product of the PDFs for the neutrino right ascension or RA ($\alpha$) and declination ($\delta$). The former is adequately approximated by a uniform distribution between 0 and $2\pi$ (i.e.
RA $\sim {\bf U}(0,\,2\pi)$), while the later is constructed using the declinations of neutrinos that are outside of the acceptance window of any SNe (See Fig. \ref{fig:bkgdir}). The acceptance window for arrival direction is determined by the 99\% lower confidence interval of Eq. \ref{eq:sigdir} (i.e., all neutrinos which satisfy $\mu_\nu \geq \mu_{99\%}$). 

%%%%%%%%%%%%%%%%%%%%%%%%%%%%%%%%%%
%%%%%%%%%%%%%%%%%%%%%%%%%%%%%%%%%%

\begin{figure}[h]
\centering
\includegraphics[width=0.45\textwidth]{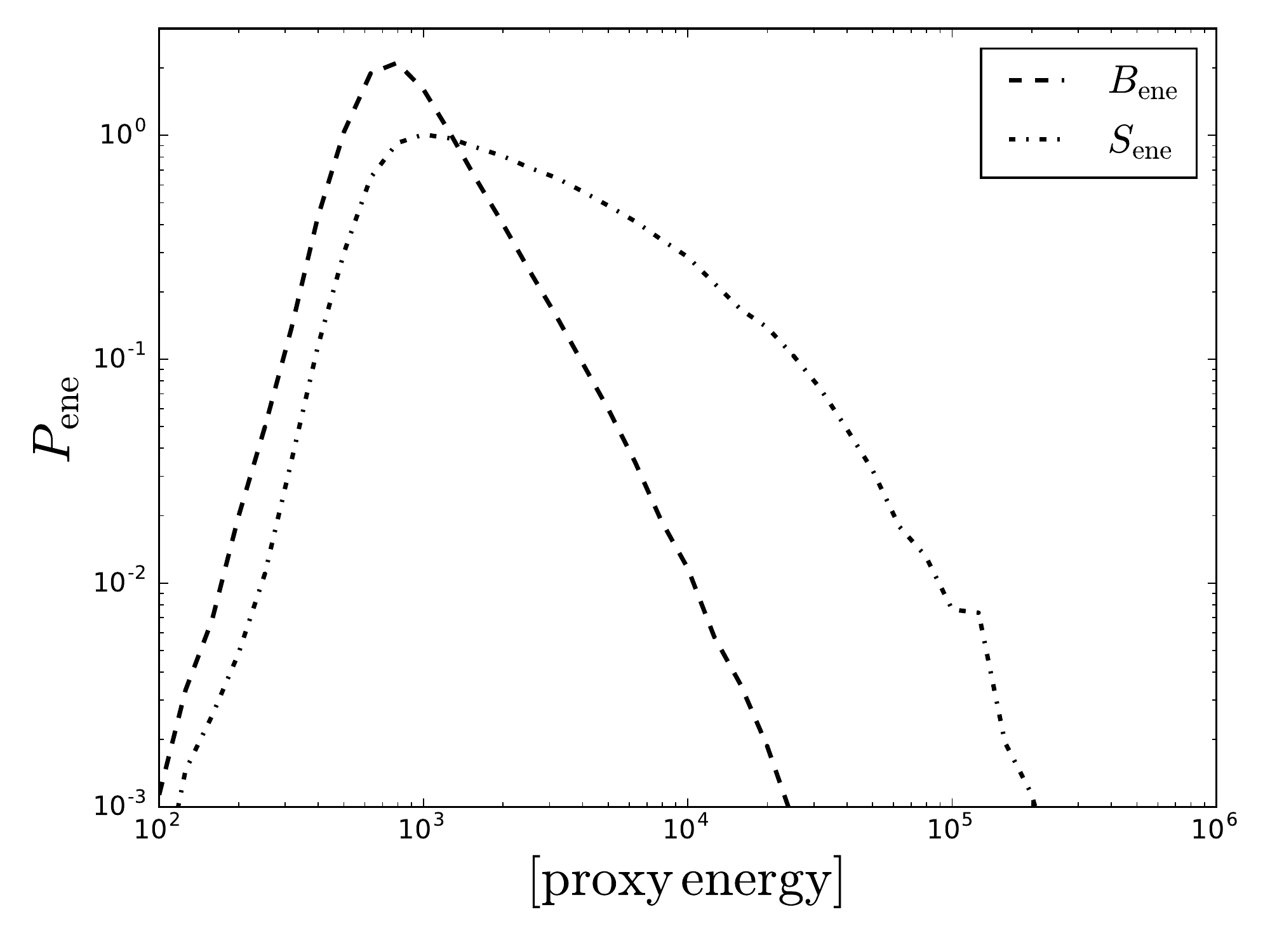}
\includegraphics[width=0.45\textwidth]{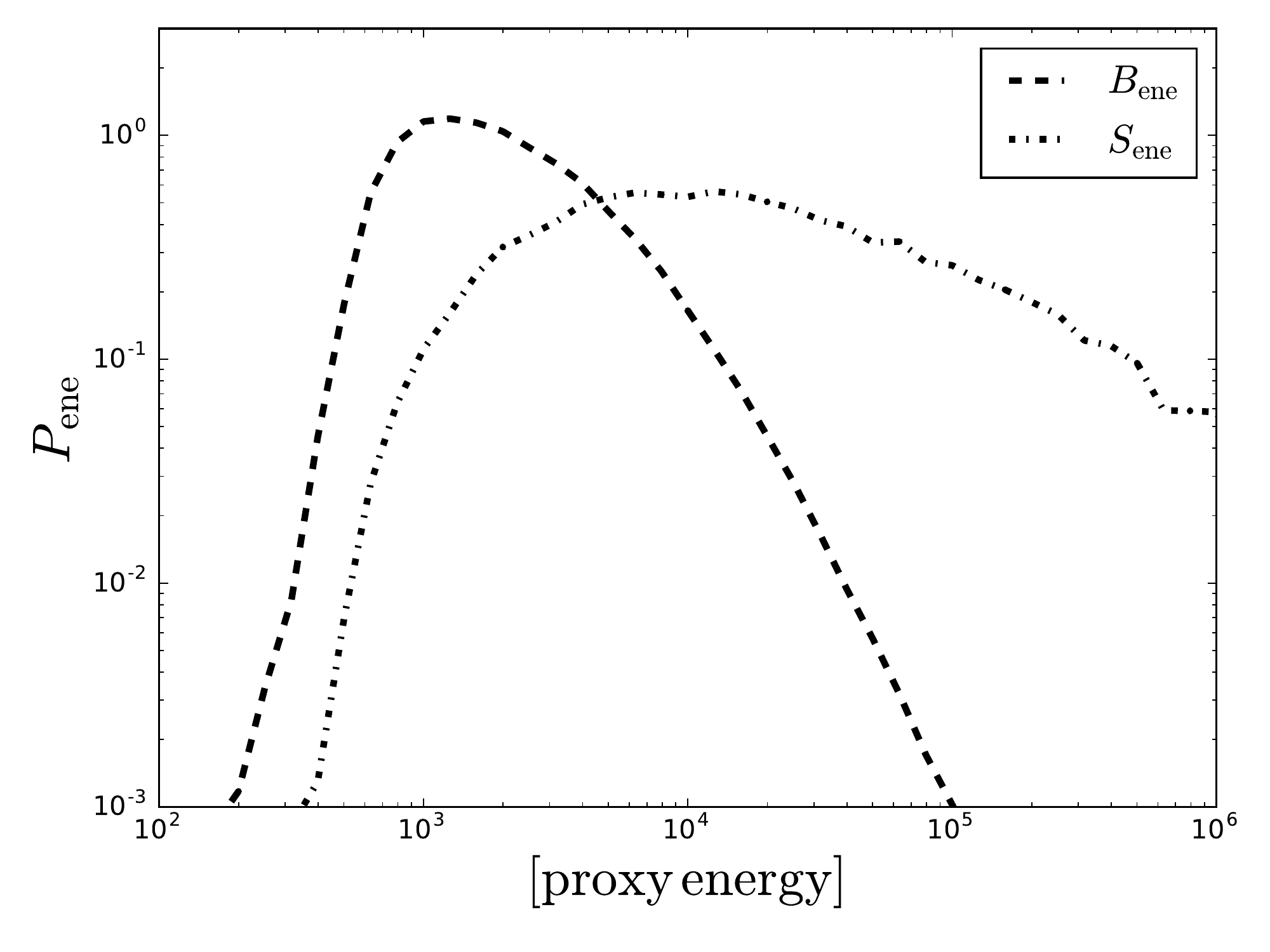}
\vspace{-1.\baselineskip}
\caption{Signal and background energy PDFs for neutrino arriving in the IceCube zenith bins $-1 \leq\cos\theta_z < -0.9$ (left) and $0.1 \leq\cos\theta_z < 0.0$ (right). Note that the IceCube detector is most sensitive to very high-energy neutrinos coming from the horizon, since the Earth is opaque to neutrinos with energy $E_\nu \gtrsim 1$ PeV. The proxy energy for each neutrino is related to the amount of photo-electrons produced in the detector.
}
\label{fig:Seng} 
\end{figure}

The energy and zenith dependent IceCube effective areas provided in Aartsen et al. (2015) \cite{Aartsen:2015de} are convolved with an unbroken $E_\nu^{-2}$ spectrum for the neutrino fluence to construct the signal PDF ($S_{\rm ene}$) for the neutrino energy proxy, which is related to the total amount of photoelectric energy a neutrino deposits in the detector (see Fig. \ref{fig:Seng}, dash-dotted lines). Later, we perform a similar analysis for a soft $E_\nu^{-2.3}$ spectrum using analytic corrections to the effective areas provided. The total amount of CR energy in the jet -- which can be related to the neutrino energy if one assumes that the choked jet is calorimetric -- is left as a tunable parameter so we can set a limit on the CR acceleration efficiency of cjSNe. The luminosity distance is obtained from the measured redshift of each SN. For the one SN for which no redshift is determined (PTF11ilr), we randomly draw a value from the remaining SN redshifts. The background energy PDF ($B_{\rm ene}$) is again constructed using neutrinos outside of the acceptance windows of all SNe, and is in good agreement with the proxy energy distribution of atmospheric neutrinos for proxy energies $\gtrsim 10^3$ (see Fig \ref{fig:Seng}, dashed lines) \cite{2010RScI...81h1101H}. Because of the geometry of the IceCube detector, and the direction and energy dependent neutrino opacity of the Earth, both the signal and background energy PDFs depend on the IceCube zenith angle  (i.e., the declination of the neutrino event, see Fig \ref{fig:Seng}). Therefore, we produce energy PDFs for each zenith bin of the IceCube effective area from Aartsen et al. 2015. 

%%%%%%%%%%%%%%%%%%%%%%%%%%%%%%%%%%
%%%%%%%%%%%%%%%%%%%%%%%%%%%%%%%%%%

\subsection{Construction of Test Statistic}

The likelihood that a given neutrino event is associated with a particular SN is related to the product of the ratio of signal(-like) and background PDFs defined above

\beq
\label{eq:singlelike}
\mathcal{\frac{S}{B}} = \frac{S_T}{B_T}\frac{S_{\rm dir}}{B_{\rm dir}}\frac{S_{\rm ene}}{B_{\rm ene}}.
\enq
The composite likelihood for each SN is the product of the likelihoods for each neutrino event associated with a SN. We perform a stacking analysis by forming a likelihood composed of the product of all SN likelihoods. Each SN likelihood is weighted by an appropriate Poisson factor, which accounts for the number of neutrinos associated with the $j$th SN as a function of its background ($b_j$) and signal ($s_j$) rates (i.e., the number of signal-like and background neutrinos found in a SN's acceptance window) 

\beq
\label{eq:poisson}
\mathcal{P}_j(s_j,b_j) = \frac{(s_j+b_j)^{N_j}}{N_j!}\,e^{-(s_j+b_j)}. 
\enq
The background rate $b_j$ is calculated using data from randomized synthetic experiments. The neutrino arrival data are scrambled to create synthetic data sets. The expected background rate for the $j$th SN is taken to be the average number of associated neutrinos from these scrambled datasets. The signal rate $s_j$ is calculated by maximizing the log-likelihood, where the likelihood function is given by ${\cal L}(\{s,b\})=\prod_{j=1}^{N_{sn}}\left[{\cal P}_j(\{s,b\})\prod_{i=1}^{N_j} {\cal L}_{i,j}(\{s,b\})\right]$. The likelihood that an individual neutrino event $i$ is associated with SN $j$ is given by 
\beq
\mathcal{L}_{i,j}(s_j,b_j)  = \frac{s_j\mathcal{S}_i+b_j\mathcal{B}_i}{s_j+b_j}.
\enq
For simplicity, we maximize the ratio of the log-likelihoods assuming a signal and no signal hypothesis, with ${\rm TS} = \ln\left[\frac{\mathcal{L}(\{s\})}{\mathcal{L}(\{0\})}\right]$. Our test statistic is then

%The former is determined using data from randomized synthetic experiments for each SN, and the later is calculated by maximizing the log-likelihood, where the likelihood function is given by $\mathcal{L}(\{s\}) = \prod_j^{N_{\rm sn}}\left[ \mathcal{P}_j\prod_{i=1}^{N_j} \mathcal{L}_i\right]$. For simplicity, we maximize the ratio of the log-likelihoods assuming a signal and no signal hypothesis, with ${\rm TS} = \ln\left[\frac{\mathcal{L}(\{s\})}{\mathcal{L}(\{0\})}\right]$. Our test statistic is then

\beq
\label{eq:ts}
{\rm TS} = \sum_j^{N_{\rm sn}} \left[-s_j + \sum_{i=1}^{N_j} \ln\left[\frac{s_j \mathcal{S}_i}{b_j \mathcal{B}_i} + 1\right]\right]. 
\enq
From Eq. \ref{eq:singlelike} ${\mathcal S}_i = S_T\,S_{\rm dir}\,S_{\rm ene}$ and likewise for ${\mathcal B}_i$. The significance of our observed test statistic ${\rm TS}_{\rm obs}$ is determined using a frequentist method. We produce synthetic background only data sets by randomizing the arrival times, directions, energy, and angular uncertainty of our original neutrino sample. We then compute a distribution of test statistics ${\rm TS_{bkg}}$ by applying Eq. \ref{eq:ts} to 100,000 of these synthetic data sets. Note that in the background only hypothesis, some neutrinos in the scrambled data are identified as signal-like neutrinos (i.e., $s_j > 0$, see the gray band of Fig. \ref{fig:s_band}). Therefore, we rejected the background only hypothesis if the observed test statistic ${\rm TS_{obs}}$ is $> 90\%$ of the distribution ${\rm TS_{bkg}}$. 

\subsection{Placing Upper Limits} 
\label{ss:upperlims}
The test statistic calculated using observed (i.e., unscrambled) data ${\rm TS_{obs}}$ is consistent with the distribution of test statistics calculated using scrambled data ${\rm TS_{bkg}}$ (See Fig. \ref{fig:t_bkg} below). This latter scenario is referred to as the background only hypothesis. Therefore, we can place upper limits on the total  SN energy deposited in CRs ${\mathcal E}_{\rm cr}$ (assuming the cjSNe are calorimetric with regards to neutrino production), and the fraction of SN which have choked jets pointed towards earth $f_{\rm jet}$. To do this, we calculate the probability that the distribution of SNe with fixed ${\mathcal E}_{\rm cr}$ and $f_{\rm jet}$ would produce a $> 90\%$ CL detection. In practice, this is given by the fractional number of ${\rm TS_{sig}}$ that are greater than 90\% of ${\rm TS_{bkg}}$. 

We accomplish this by injecting true signal neutrinos into our randomized synthetic data sets. For each data set, $f_{\rm jet}\,N_{\rm sn}$ SNe are chosen from random to produce signal neutrinos. These signal neutrinos have arrival time and direction, as well as energies drawn from the signal PDFs described above. The number of true signal neutrinos injected in the acceptance window of each SN is determined using the expected fluence from that SN. The muon neutrino fluence per logarithmic energy interval from a calorimetric SN jet measured on Earth (i.e., after flavor mixing) can be approximated as \cite{1999PhRvD..59b3002W}

\beq
\label{eq:fluence}
\mathcal{F}_{\nu_\mu} \approx \frac{1}{8}\frac{{\mathcal E}_{\rm cr}}{4\pi D_L^2\mathcal{R}_{\rm cr}}\;{\rm erg\,cm^{-2}},
\enq
where $D_L$ is the luminosity distance of the SN, and $\mathcal{R}_{\rm cr} = \ln(\varepsilon_{\rm cr,max}/\varepsilon_{\rm cr,min})\simeq 18$ is a bolometric correction factor for a $n_{\varepsilon_{\rm cr}} \propto \varepsilon_{\rm cr}^{-2}$ CR spectrum. Note that the factor of $\frac{1}{8}$ is related to the branching ratios of neutrino and charged pions, the later of which decay to produce neutrinos (see e.g., Waxman \& Bahcall 1998 for a review \cite{1999PhRvD..59b3002W}). The number of signal neutrinos is drawn from a Poisson distribution, with mean given by the expected number of signal neutrinos. The latter is determined by the product of the SN neutrino fluence with IceCube's effective area. While we use the neutrino effective area given by Ref.~\cite{Aartsen:2014cva}, the energy and zenith-angle averaged value is approximately 
%$\tilde{\mathcal A}_{\rm eff}\sim 10^{4}\,{\rm cm^2\,erg^{-1}}$. 
$\bar{\mathcal A}_{\rm eff}\sim 10^{4}\,{\rm cm^2}$.
\section{Results}

Based on the distribution of ${\rm TS}$ assuming a background only model (i.e., scrambling the existing neutrino data without injecting signals), we do not see any significant association between the HE IceCube neutrinos and optically detected Type Ibc SNe for the time period May 2011-May 2012. 
%%%%%%%%%%%%%%%%%%%%%%%%%%%%%%%%%%
\begin{figure}[H]
\centering
\includegraphics[width=0.8\textwidth,height=3.5in]{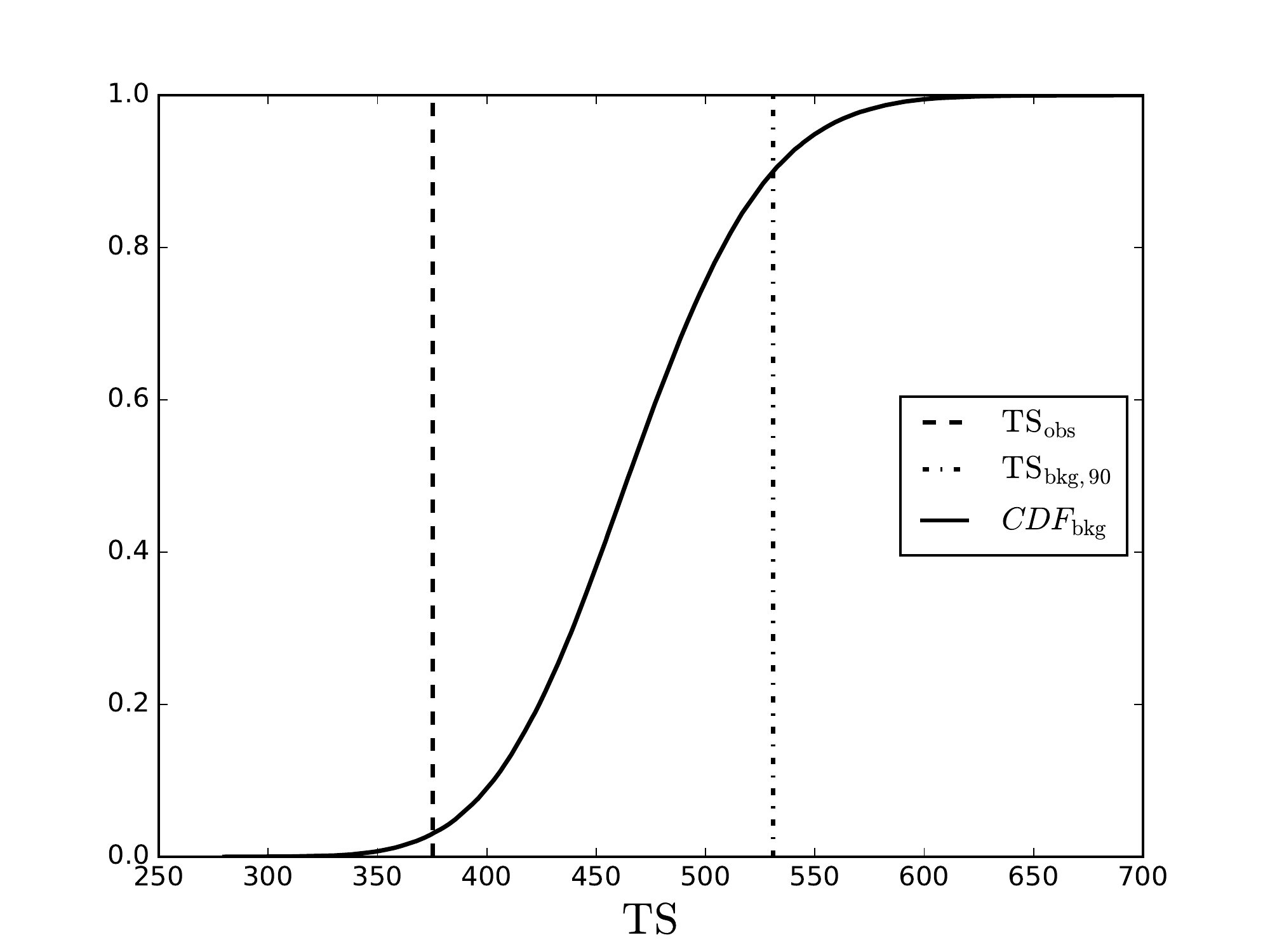}
\vspace{-1.\baselineskip}
\caption{Cumulative distribution function of ${\rm TS}_{\rm bkg}$ generated from synthetic experiments of randomized neutrino data with the observed test statistic ${\rm TS}_{\rm obs}$ (dashed line) and the 90\% upper limit on the background distribution ${\rm TS}_{\rm bkg, 90}$ used to set exclusion contours for the signal hypothesis (dash-dot, see text for details). ${\rm TS}_{\rm obs}$ is consistent with a background only hypothesis. 
}
\label{fig:t_bkg} 
\end{figure}
%%%%%%%%%%%%%%%%%%%%%%%%%%%%%%%%%%

Fig \ref{fig:t_bkg} gives the cumulative distribution function (CDF) of ${\rm TS}$ assuming a background only hypothesis, with ${\rm TS}_{\rm obs}$ given by the dashed line. Each SN was associated with $\sim 15$ track events on average. This level of association was expected from the synthetic experiments. 
Large values of ${\rm TS}$ correspond to experiments with many signal-like neutrinos, while values of ${\rm TS}\approx0$ indicate few signal-like neutrinos. 

\begin{figure}[H] 
\centering
\includegraphics[width=0.8\textwidth,height=3.0in]{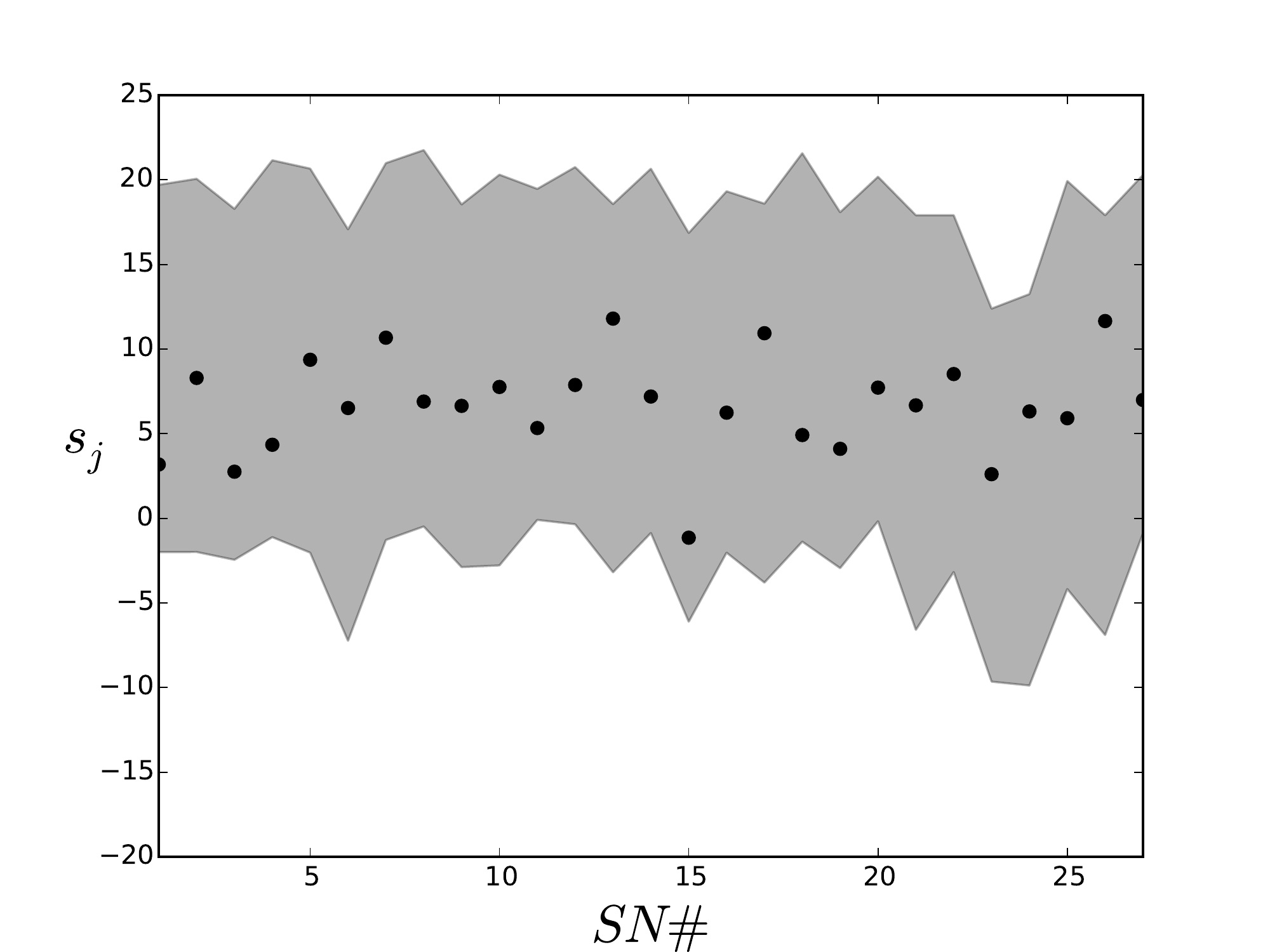}
\vspace{-1.\baselineskip}
\caption{Best fit value for the number of signal-like neutrinos $(s_j)$ for each SN in our sample (with their label number on the x-axis). The black dots correspond to the best fit values from our observed data, while the gray band is the 99\% confidence interval determined from the synthetic data samples of randomized data. Note that $s_j \neq 0$ for all cases, meaning that some neutrinos from the scrambled data sets are identified as signal-like neutrinos. We do not see a significant number of signal-like neutrinos above what is expected from a background only model. 
}
\label{fig:s_band}
\end{figure}
Looking at the best fit values for the number of signal-like neutrinos in our observed data, we find that they are within the 99\% confidence interval determined by the synthetic background data distribution (See Fig \ref{fig:s_band}). This gives a further indication that we do not see a significant association between HE neutrinos and the 29 Type Ibc SNe in our sample. 

\begin{figure}[h]
\centering
\includegraphics[width=0.45\textwidth]{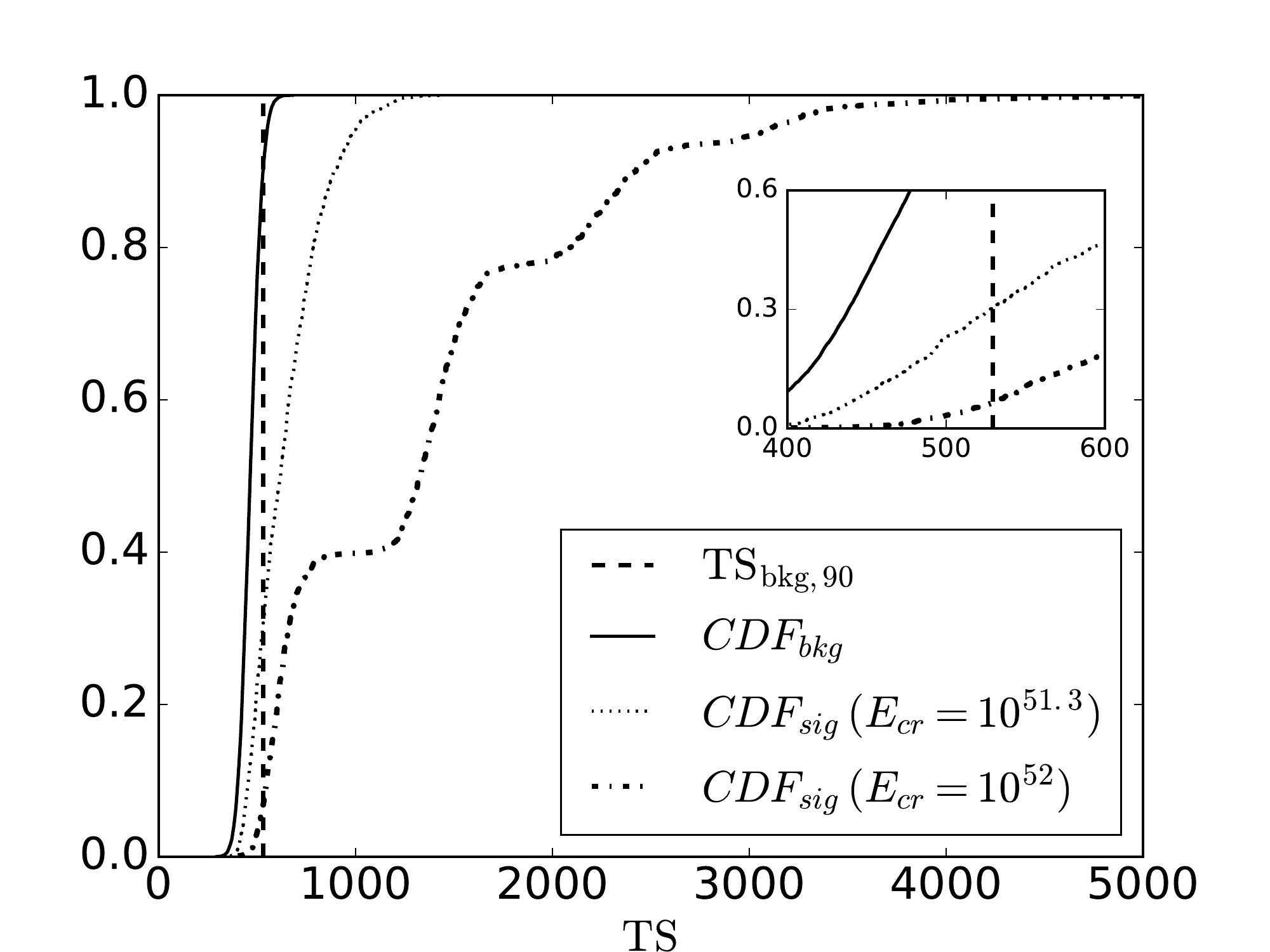}
\includegraphics[width=0.45\textwidth]{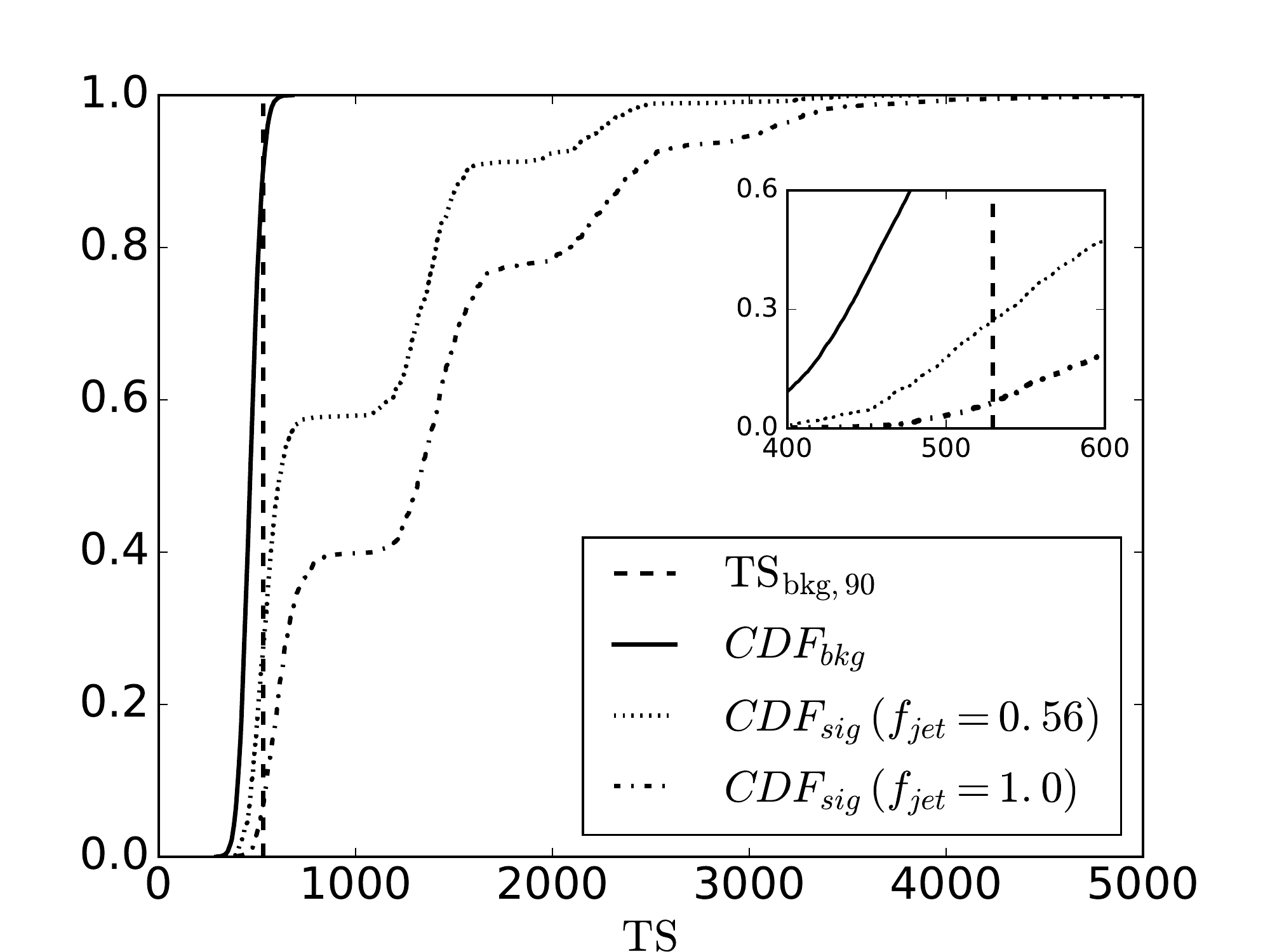}
\vspace{-1.\baselineskip}
\caption{Cumulative distribution functions of ${\rm TS}$ comparing the background and signal scenarios with ${\rm TS}_{\rm bkg, 90}$. We show how the CDF produced by the signal hypothesis changes as $\mathcal{E}_{\rm cr}$ is varied with $f_{\rm jet} = 1.0$ ({\bf left}), and for different values of $f_{\rm jet}$ with ${\mathcal E}_{\rm cr} = 10^{52}\,{\rm erg}$ ({\bf right}). Insets show where ${\rm TS}_{\rm bkg,90}$ intersects with each CDF. The wavy features of the signal CDFs are a results of SNe from different luminosity distances being included or excluded in a synthetic experiment. When ${\mathcal E}_{\rm cr}$ is the same for all SNe, closer jetted SNe produce a higher fluence of signal neutrinos in the detector. 
}
\label{fig:compare} 
\end{figure}

We also produce distributions of ${\rm TS}_{\rm sig}$, where true signal neutrinos are injected into our sample. Fig. \ref{fig:compare} (left) compares the CDF of the background only scenario (solid line) with different distributions of ${\rm TS}_{\rm sig}$ with fixed $f_{\rm jet} = 1.0$ and ${\mathcal E}_{\rm cr} = 10^{51.3}\,{\rm erg}$ (dotted), ${\mathcal E}_{\rm cr} = 10^{52} \,{\rm erg}$ (dash-dotted). Fig. \ref{fig:compare} (right) compares the background only CDF with two different signal distributions, with fixed ${\mathcal E}_{\rm cr} = 10^{52}\,{\rm erg}$ and $f_{\rm jet}$ = 0.56 (dashed) and $f_{\rm jet} = 1$ (dash-dotted). These distributions are compared with the observed value ${\rm TS}_{\rm bkg,90}$ (thick dashed) to determine what is the probability of making a 90\% CL observation for a particular signal hypothesis. The wavy features for the signal CDFs are a result of SNe from varying luminosity distances being included or excluded in a synthetic experiment. When ${\mathcal E}_{\rm cr}$ is the same for all SNe, closer jetted SNe result a higher fluence of signal neutrinos in the detector. Note that, as the number of signal neutrinos in our synthetic data set decrease -- as either ${\mathcal E}_{\rm cr}$ or $f_{\rm jet}$ decrease --  the signal distribution of ${\rm TS}$ begins to resemble that of the background only scenario. 

\begin{figure}[H]
\centering
\includegraphics[width=0.8\textwidth]{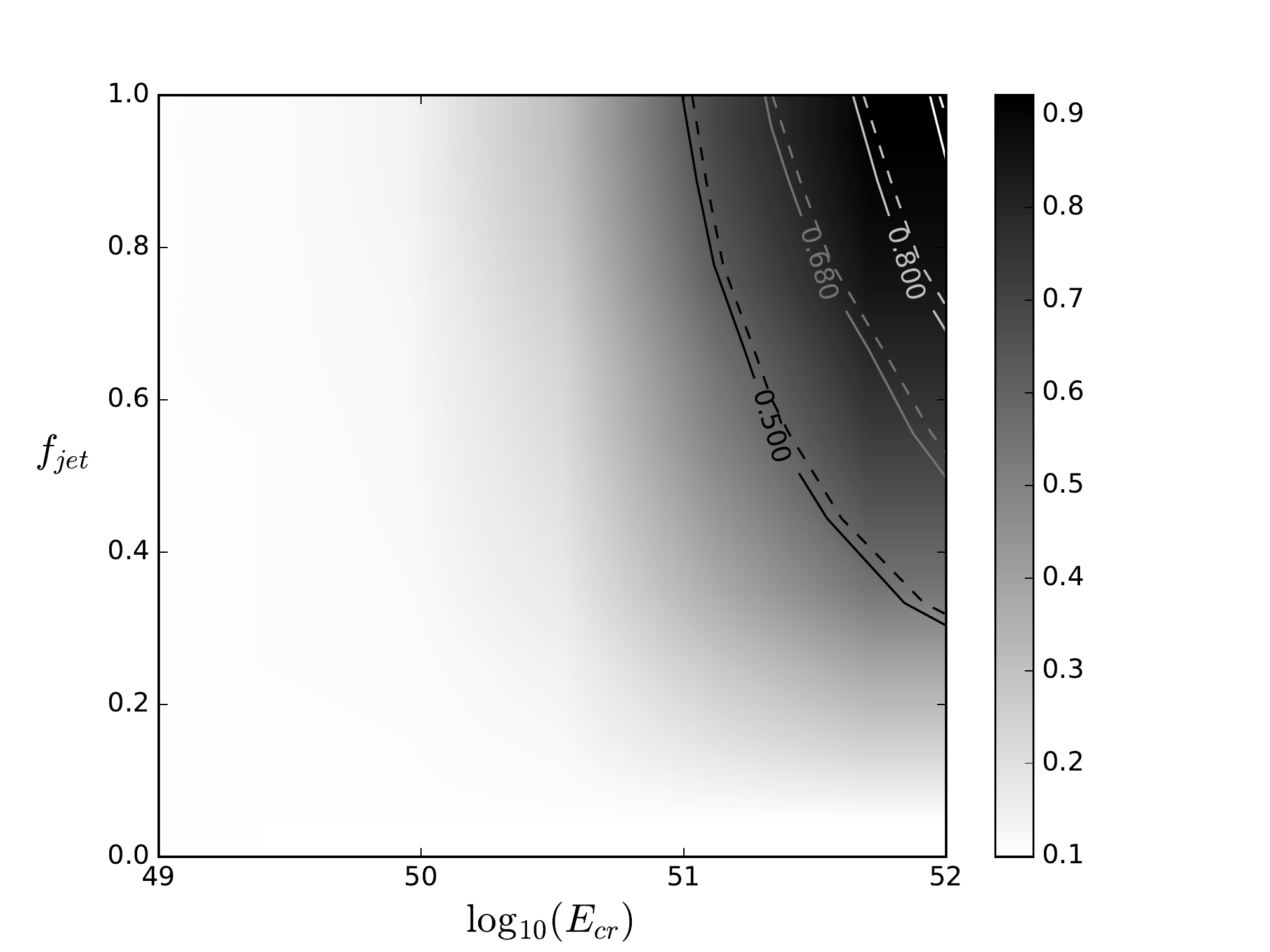}
\vspace{-1.\baselineskip}
\caption{Heat map that shows the exclusion limits for values of $\mathcal{E}_{\rm cr}$ and $f_{\rm jet}$ based on synthetic experiments with true signal neutrinos added. Darker color means higher confidence in exclusion (e.g., the top right corner is excluded at 90\% confidence). Solid lines shows the exclusion region for an $E_\nu^{-2}$ neutrino spectrum, while the dashed lines are for an $E_\nu^{-2.3}$ spectrum. 
}
\label{fig:excl} 
\end{figure}

We are able to place upper limits on the fraction of SNe which have choked jets pointed towards us $f_{\rm jet}$, and the total amount of CR energy contained in such jets ${\mathcal E}_{\rm cr}$. Fig \ref{fig:excl} shows a heat map of the exclusion region in the ${\mathcal E}_{\rm cr}-f_{\rm jet}$ plane which is given by the probability of observing $ {\rm TS_{sig}} > 90\%$ of ${\rm TS_{bkg}}$  (see \S \ref{ss:upperlims}). We consider the values $0 \leq f_{\rm jet} \leq 1$ to consider the scenario where no Type Ibc SNe have jets, and when all such SNe contain them. The solid (dashed) lines show the exclusion contours for a $E_\nu^{-2}$ ($E_\nu^{-2.3}$) neutrino spectrum. We are able to place limits on $f_{\rm jet}$ down to ${\mathcal E}_{\rm cr}\sim 10^{51.5}\,{\rm erg}$, which is comparable to a typical SN explosion energy. Note that ${\mathcal E}_{\rm cr}$ is the isotropic equivalent energy, so the true amount of jet energy contained in CRs would be reduced by a factor $\theta^2_j/2\sim 10^{-1}$ depending on the opening angle of the choked jet $\theta_j$. 

We can compare our result with a simple analytic argument, using the typical IceCube fluence sensitivity $\phi_{\rm lim}\sim 10^{-4}\,{\rm erg\,cm^{-2}}$ and summing over each SN. The limit on ${\mathcal E}_{\rm cr}$ and $f_{\rm jet}$ is set using the Poissonian probability of observing more neutrinos than the 90\% upper limit assuming a background only hypothesis ($N_{\rm bkg,90}$). With the total neutrino background rate (i.e., $n_b = \sum_{j=1}^{N_{\rm sn}} b_j$) the 90\% upper limit on the number of observed neutrinos in the background only hypothesis is calculated solving \cite{1986ApJ...303..336G}

\beq
\label{eq:bkg_90}
\sum_{x=0}^{N_{\rm bkg,90}}  \frac{n_b^x\,e^{-n_b}}{x!} \leq 0.1 
\enq
The probability $\mathcal{P}_{> 90}$ of observing more neutrinos given a signal rate ($n_s$), assuming the average isotropic equivalent CR energy released per burst is $\tilde{\mathcal E}_{\rm cr} = {\mathcal E}_{\rm cr}\,f_{\rm jet}$, is given by 

\beq
\label{eq:90}
\mathcal{P}_{>90} = \sum_{y = 0}^{N_{\rm bkg,90}}\frac{(n_s+n_b)^y\,e^{-(n_s+n_b)}}{y!},
\enq
where $n_s$ is estimated to be
\beq
\label{eq:sig_90}
n_s = {\phi}_{\rm lim}^{-1} \frac{\tilde{\mathcal E}_{\rm cr}}{32\pi\mathcal{C}}\,\sum_{j=1}^{N_{\rm sn}} \frac{1}{D_{L,j}^2}, 
\enq
For the 28 SNe in our sample with a measured redshift, this gives $\tilde{\mathcal E}_{\rm cr,90\%} \sim 10^{52}\,{\rm erg}$.  

\begin{figure}[H]
\centering
\includegraphics[width=0.8\textwidth]{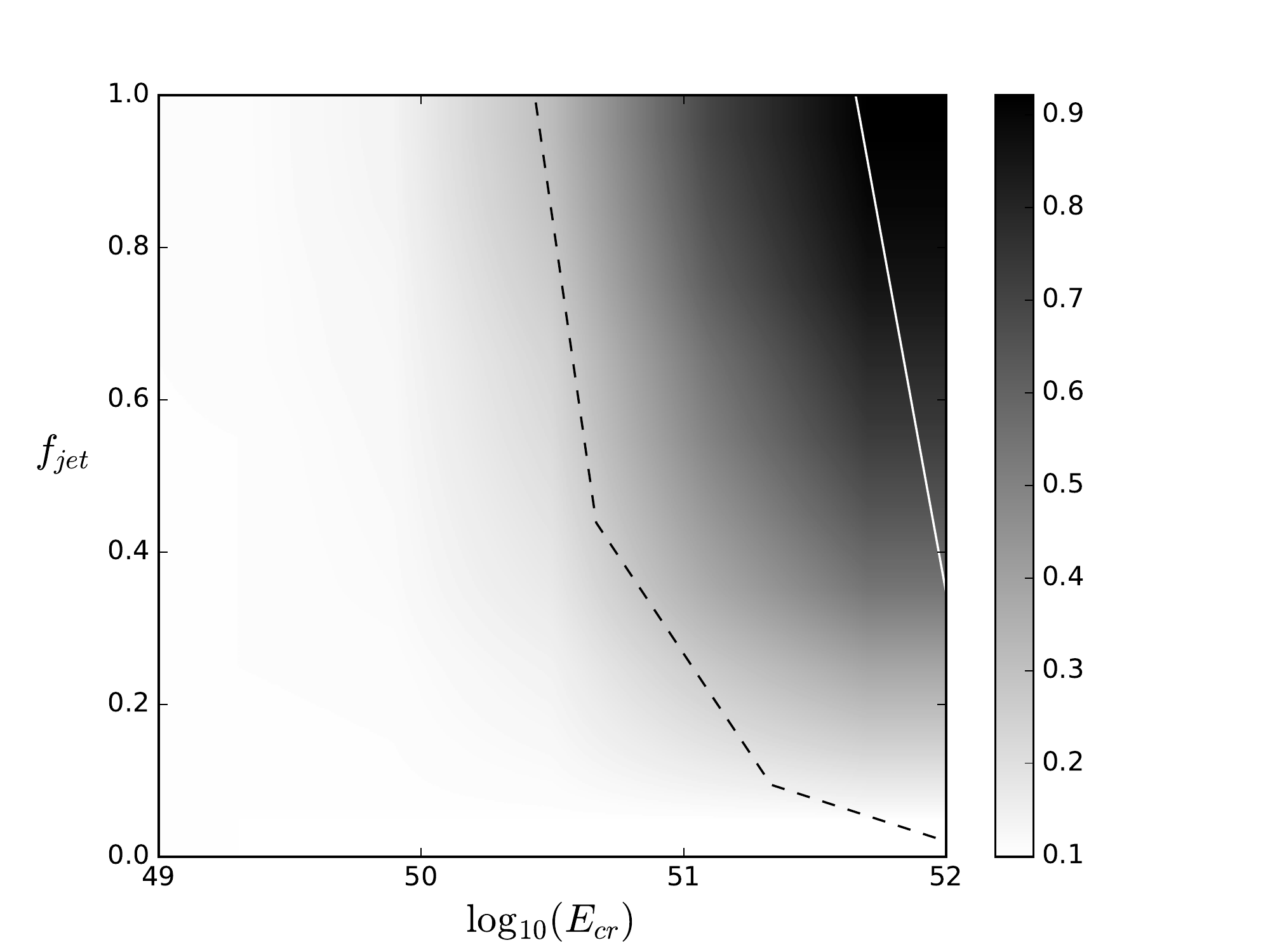}
\vspace{-1.\baselineskip}
\caption{Comparison of the numerical results of our analysis (heat map, see Fig. \ref{fig:excl}) with the 90\% upper limit determined by Eq. \ref{eq:90}. The white solid line corresponds to the 90\% upper limit using the 28 SNe with measured redshift, while the black dashed line corresponds to the analytic 90\% confidence level using Eq. \ref{eq:90} and 131 SNe from May 2010-May 2017 (see text for details).
}
\label{fig:excl_analytic} 
\end{figure}

Fig. \ref{fig:excl_analytic} compares the heat map of our numerical results (as seen in Fig. \ref{fig:excl}) with the analytic results produced by Eq. \ref{eq:90} (white solid line). We see that there is reasonable agreement between the shape of the exclusion region from both methods, as well as the location of the 90\% confidence limit at ${\mathcal E}_{\rm cr} \sim 10^{52}\,{\rm erg}$ (for $f_{\rm jet} = 1$). With an additional 6 years of IceCube data, we find using Eqs. \ref{eq:bkg_90}-\ref{eq:90} that the 90\% confidence limit on ${\mathcal E}_{\rm cr}$ can be improved by a factor of $\sim  10$  (see Fig. \ref{fig:excl_analytic} black dashed line, which was calculated using 131 Type Ibc SNe that were observed between May 2010-May 2017 with an extrapolation of the expected neutrino background rate for 7 years of data from 1 year of data). 

\section{Summary and Discussion}

We performed an unbinned log-likelihood analysis of the $\sim 70,000$ up-going, track-like neutrino events observed by IceCube between May 2011 and May 2012 and 29 Type Ibc SNe observed during this same time period. We found no significant excess of signal neutrinos associated with these SNe (see Fig. \ref{fig:t_bkg}). Based on this non-detection, we were able to place upper limits on the fraction of Type Ibc SNe which may harbor a choked jet pointed towards Earth, and the total amount of CR energy they can produce assuming such jets are efficient neutrino factories. 
Our upper limits will be improved by a factor of $\sim 10$ once the remaining 6 years of IceCube data are made public. 

If cjSNe significantly contribute to the observed IceCube flux below 100~TeV, the required CR energy production rate is $3\times{10}^{53}~{\rm erg}~{\rm Mpc}^{-3}~{\rm yr}^{-1}$. For Type Ibc SNe with density $2\times{10}^{4}~{\rm Gpc}^{-3}$, the required CR energy per event is ${\mathcal E}_{\rm cr}\sim10^{49}$~erg for $f_{\rm jet}=1$. In reality, $f_{\rm jet}\ll 1$ is expected, so the number of ``nearby'' SN samples should be large enough ($\gg 1/f_{\rm jet}$) to obtain meaningful constraints and test the choked jet scenario for IceCube neutrinos~\cite{Murase:2013ffa}. IceCube currently has upgoing track-like data for $\sim 7$ years (2010-2017). When this data is made available to the public, combined with the increased detection rate of ccSNe with ASAS-SN and other all-sky optical surveys -- there are an additional 104 Type Ibc SNe that were detected from May 2012 to May 2017 -- cjSNe can be more tightly tested as potential sources of HE astrophysical neutrinos in future, especially with IceCube-Gen2.

Non-detection of neutrinos from cjSNe may not be surprising at all. Theoretically, the standard Fermi acceleration mechanism is inhibited in radiation mediated shocks, so powerful jets or compact progenitors would not be ideal as high-energy neutrino emitters via this process~\cite{Murase:2013ffa}. This also implies that one has to be careful to use the neutrino data to constrain physical quantities of choked jets, except for the CR acceleration efficiency. 
On the other hand, a correlation between HE neutrinos and cjSNe would indicate that high-energy CRs can be accelerated in relativistic jets launched during stellar collapse, which has important implications for CR acceleration in dense environments. Sub-TeV neutrino signals may allow us to probe other alternative acceleration mechanisms, such as the neutron-proton-converter acceleration process~\cite{Murase:2013hh,Kashiyama:2013ata}. Also, a new window will have been opened into the interior of old, massive stars. Except for the thermal neutrinos observed from SN1987A, these high energy neutrinos would constitute the first direct observation of an exploding star beneath its photosphere\footnote{the region at which the stellar material becomes optically thick}. Such observations could allow stellar interior models to be directly tested using methods such as ``neutrino tomography'' \cite{Razzaque:2003eq}.

Our analysis is one of the first systematic searches for neutrino sources from a population of choked jet SNe. 
We have assumed a $E_\nu^{-2}$ spectrum. While it is reasonable to assume the neutrino energy spectra is given approximately flat, these spectra are generally soft at high energies $E_\nu\sim 0.1-1\,{\rm PeV}$, and peak in the energy range that IceCube is most sensitive to $E_\nu\sim 100\,{\rm TeV}$. Using an approximation of the IceCube effective area for a soft, unbroken neutrino spectrum ($E^{-2.3}$), our results change by less than 10\%. 
Future works will be able to improve the accuracy of this analysis by constructing an energy signal PDF by convolving different cjSNe neutrino models with the exact IceCube effective area. 

Searches that rely on optical data only, or with some combination of soft $\gamma$-rays/X-rays are becoming increasingly important, as $\gamma$-ray bright sources such as GRBs and blazars are being ruled out as the main sources of the IceCube neutrinos. Furthermore, current and soon to be operational optical surveys will view the entire optical sky within 3 days up to a depth of 20.4 mag. 
By constantly monitoring the sky, we will gain a better understanding of the early explosion processes of ccSNe (i.e., $\lesssim 1$ day after the initial explosion). The upgraded Zwicky Transient Facility ``will detect one SN within 24 hours of its explosion {\it every night}"\cite{Bellm:2014tu}.  This benefits the analysis in two ways: 1.) It allows for a more accurate determination of the time window during which a choked jet may be formed in the ccSNe, reducing the signal time window from $\sim 15$ days, to hours. 
2.) Early observations of SNe explosions will give the first direct observations of the material immediately surrounding a SN progenitor. Currently the mass loss processes which occur during the last $0.1-10$ year of a stripped-envelope progenitor are not well understood, and could provide insight into which types of stars will form a jet, and ultimately a typical Type Ibc SNe, a cjSNe, or a full GRB. This in turn will help guide searches for the most likely transients to produce HE neutrinos.

%%%%%%%%%%%%%%%%%%%%%%%%%%%%%%%%%%%%%%%%%%%%%%%%%%
%%%%%%%%%%%%%%%%%%%%%%%%%%%%%%%%%%%%%%%%%%%%%%%%%%

%\medskip
\acknowledgments
We are grateful for partial support from NASA NNX13AH50G (N.S. and P.M.) and
NSF PHY-1620777 (K.M.). We would like to thank Derek Fox and Anna Franckowiak for useful discussions. 

%%%%%%%%%%%%%%%%%%%%%%%%%%%%%%%%%%%%%%%%%%%%%%%%%%
%%%%%%%%%%%%%%%%%%%%%%%%%%%%%%%%%%%%%%%%%%%%%%%%%%
\bibliographystyle{revtex}
\bibliography{sne17}

\hyphenation{Post-Script Sprin-ger}
\begin{thebibliography}{10}
\providecommand*{\bibinfo}[2]{#2}
\providecommand*{\eprint}[1]{#1}
\providecommand*{\url}[1]{#1}
\bibitem{2013PhRvL.111b1103A}
\bibinfo{author}{M.~G. Aartsen}, \bibinfo{author}{R.~Abbasi},
  \bibinfo{author}{Y.~Abdou}, \bibinfo{author}{M.~Ackermann},
  \bibinfo{author}{J.~Adams}, \bibinfo{author}{J.~A. Aguilar},
  \bibinfo{author}{M.~Ahlers}, \bibinfo{author}{D.~Altmann},
  \bibinfo{author}{J.~Auffenberg}, \bibinfo{author}{X.~Bai}, \emph{et~al.},
  \bibinfo{journal}{Phys. Rev. Lett.} \bibinfo{volume}{\textbf{111}}(2),
  \bibinfo{pages}{021103} (\bibinfo{date}{Jul. 2013}).
\bibitem{Aartsen:2013jdh}
\bibinfo{author}{M.~Aartsen} \emph{et~al.} (\bibinfo{collaboration}{IceCube
  Collaboration}), \bibinfo{journal}{Science} \bibinfo{volume}{\textbf{342}},
  \bibinfo{pages}{1242856} (\bibinfo{date}{2013}), \eprint{1311.5238}.
\bibitem{2014PhRvL.113j1101A}
\bibinfo{author}{M.~G. Aartsen}, \bibinfo{author}{M.~Ackermann},
  \bibinfo{author}{J.~Adams}, \bibinfo{author}{J.~A. Aguilar},
  \bibinfo{author}{M.~Ahlers}, \bibinfo{author}{M.~Ahrens},
  \bibinfo{author}{D.~Altmann}, \bibinfo{author}{T.~Anderson},
  \bibinfo{author}{C.~Arguelles}, \bibinfo{author}{T.~C. Arlen}, \emph{et~al.},
  \bibinfo{journal}{Phys. Rev. Lett.} \bibinfo{volume}{\textbf{113}}(1),
  \bibinfo{pages}{101101} (\bibinfo{date}{Sep. 2014}).
\bibitem{2015PhRvD..91b2001A}
\bibinfo{author}{M.~G. Aartsen}, \bibinfo{author}{M.~Ackermann},
  \bibinfo{author}{J.~Adams}, \bibinfo{author}{J.~A. Aguilar},
  \bibinfo{author}{M.~Ahlers}, \bibinfo{author}{M.~Ahrens},
  \bibinfo{author}{D.~Altmann}, \bibinfo{author}{T.~Anderson},
  \bibinfo{author}{C.~Arguelles}, \bibinfo{author}{T.~C. Arlen}, \emph{et~al.},
  \bibinfo{journal}{Phys. Rev. D} \bibinfo{volume}{\textbf{91}}(2),
  \bibinfo{pages}{022001} (\bibinfo{date}{Jan. 2015}).
\bibitem{Aartsen:2015de}
\bibinfo{author}{M.~G. Aartsen}, \bibinfo{author}{K.~Abraham},
  \bibinfo{author}{M.~Ackermann}, \bibinfo{author}{J.~Adams},
  \bibinfo{author}{J.~A. Aguilar}, \bibinfo{author}{M.~Ahlers},
  \bibinfo{author}{M.~Ahrens}, \bibinfo{author}{D.~Altmann},
  \bibinfo{author}{T.~Anderson}, \bibinfo{author}{M.~Archinger}, \emph{et~al.},
  \bibinfo{journal}{Phys. Rev. Lett.} \bibinfo{volume}{\textbf{115}}(8),
  \bibinfo{pages}{081102} (\bibinfo{date}{Aug. 2015}).
\bibitem{Aartsen:2016xlq}
\bibinfo{author}{M.~G. Aartsen} \emph{et~al.} (\bibinfo{collaboration}{IceCube
  Collaboration}), \bibinfo{journal}{Astrophys. J.}
  \bibinfo{volume}{\textbf{833}}(1), \bibinfo{pages}{3} (\bibinfo{date}{2016}),
  \eprint{1607.08006}.
\bibitem{2015ApJ...809...98A}
\bibinfo{author}{M.~G. Aartsen}, \bibinfo{author}{K.~Abraham},
  \bibinfo{author}{M.~Ackermann}, \bibinfo{author}{J.~Adams},
  \bibinfo{author}{J.~A. Aguilar}, \bibinfo{author}{M.~Ahlers},
  \bibinfo{author}{M.~Ahrens}, \bibinfo{author}{D.~Altmann},
  \bibinfo{author}{T.~Anderson}, \bibinfo{author}{M.~Archinger}, \emph{et~al.},
  \bibinfo{journal}{ApJ} \bibinfo{volume}{\textbf{809}}(1), \bibinfo{pages}{98}
  (\bibinfo{date}{Aug. 2015}).
\bibitem{1997PhRvL..78.2292W}
\bibinfo{author}{E.~Waxman} and \bibinfo{author}{J.~Bahcall},
  \bibinfo{journal}{Phys. Rev. Lett.} \bibinfo{volume}{\textbf{78}}(1),
  \bibinfo{pages}{2292} (\bibinfo{date}{Mar. 1997}).
\bibitem{Murase:2005hy}
\bibinfo{author}{K.~Murase} and \bibinfo{author}{S.~Nagataki},
  \bibinfo{journal}{Phys. Rev.} \bibinfo{volume}{\textbf{D73}},
  \bibinfo{pages}{063002} (\bibinfo{date}{2006}), \eprint{astro-ph/0512275}.
\bibitem{Petropoulou:2014lja}
\bibinfo{author}{M.~Petropoulou}, \bibinfo{author}{D.~Giannios}, and
  \bibinfo{author}{S.~Dimitrakoudis}, \bibinfo{journal}{Mon. Not. Roy. Astron.
  Soc.} \bibinfo{volume}{\textbf{445}}(1), \bibinfo{pages}{570}
  (\bibinfo{date}{2014}), \eprint{1405.2091}.
\bibitem{Bustamante:2015fb}
\bibinfo{author}{M.~Bustamante}, \bibinfo{author}{P.~Baerwald},
  \bibinfo{author}{K.~Murase}, and \bibinfo{author}{W.~Winter},
  \bibinfo{journal}{Nature Communications} \bibinfo{volume}{\textbf{6}},
  \bibinfo{pages}{6783} (\bibinfo{date}{Apr. 2015}).
\bibitem{Bustamante:2016wpu}
\bibinfo{author}{M.~Bustamante}, \bibinfo{author}{K.~Murase},
  \bibinfo{author}{W.~Winter}, and \bibinfo{author}{J.~Heinze},
  \bibinfo{journal}{Astrophys. J.} \bibinfo{volume}{\textbf{837}}(1),
  \bibinfo{pages}{33} (\bibinfo{date}{2017}), \eprint{1606.02325}.
\bibitem{1995APh.....3..295M}
\bibinfo{author}{K.~Mannheim}, \bibinfo{journal}{Astroparticle Physics}
  \bibinfo{volume}{\textbf{3}}(3), \bibinfo{pages}{295} (\bibinfo{date}{May
  1995}).
\bibitem{Atoyan:2001ey}
\bibinfo{author}{A.~Atoyan} and \bibinfo{author}{C.~D. Dermer},
  \bibinfo{journal}{Phys.Rev.Lett.} \bibinfo{volume}{\textbf{87}},
  \bibinfo{pages}{221102} (\bibinfo{date}{2001}), \eprint{astro-ph/0108053}.
\bibitem{Murase:2014foa}
\bibinfo{author}{K.~Murase}, \bibinfo{author}{Y.~Inoue}, and
  \bibinfo{author}{C.~D. Dermer}, \bibinfo{journal}{Phys.Rev.}
  \bibinfo{volume}{\textbf{D90}}, \bibinfo{pages}{023007}
  (\bibinfo{date}{2014}), \eprint{1403.4089}.
\bibitem{Dermer:2014vaa}
\bibinfo{author}{C.~D. Dermer}, \bibinfo{author}{K.~Murase}, and
  \bibinfo{author}{Y.~Inoue}, \bibinfo{journal}{JHEAp}
  \bibinfo{volume}{\textbf{3-4}}, \bibinfo{pages}{29} (\bibinfo{date}{2014}),
  \eprint{1406.2633}.
\bibitem{Tavecchio:2014xha}
\bibinfo{author}{F.~Tavecchio}, \bibinfo{author}{G.~Ghisellini}, and
  \bibinfo{author}{D.~Guetta}, \bibinfo{journal}{Astrophys.J.}
  \bibinfo{volume}{\textbf{793}}, \bibinfo{pages}{L18} (\bibinfo{date}{2014}).
\bibitem{Padovani:2015mba}
\bibinfo{author}{P.~Padovani}, \bibinfo{author}{M.~Petropoulou},
  \bibinfo{author}{P.~Giommi}, and \bibinfo{author}{E.~Resconi},
  \bibinfo{journal}{Mon.Not.Roy.Astron.Soc.} \bibinfo{volume}{\textbf{452}}(2),
  \bibinfo{pages}{1877} (\bibinfo{date}{2015}), \eprint{1506.09135}.
\bibitem{Aartsen:2014aqy}
\bibinfo{author}{M.~Aartsen} \emph{et~al.} (\bibinfo{collaboration}{IceCube
  Collaboration}), \bibinfo{journal}{Astrophys.J.}
  \bibinfo{volume}{\textbf{805}}(1), \bibinfo{pages}{L5}
  (\bibinfo{date}{2015}), \eprint{1412.6510}.
\bibitem{Murase:2016gly}
\bibinfo{author}{K.~Murase} and \bibinfo{author}{E.~Waxman},
  \bibinfo{journal}{Phys. Rev.} \bibinfo{volume}{\textbf{D94}}(10),
  \bibinfo{pages}{103006} (\bibinfo{date}{2016}), \eprint{1607.01601}.
\bibitem{Aartsen:2016lir}
\bibinfo{author}{M.~G. Aartsen} \emph{et~al.}
  (\bibinfo{collaboration}{IceCube}), \bibinfo{journal}{Astrophys. J.}
  \bibinfo{volume}{\textbf{835}}(1), \bibinfo{pages}{45}
  (\bibinfo{date}{2017}), \eprint{1611.03874}.
\bibitem{Murase:2016ck}
\bibinfo{author}{K.~Murase}, \bibinfo{author}{D.~Guetta}, and
  \bibinfo{author}{M.~Ahlers}, \bibinfo{journal}{Phys. Rev. Lett.}
  \bibinfo{volume}{\textbf{116}}(7), \bibinfo{pages}{071101}
  (\bibinfo{date}{Feb. 2016}).
\bibitem{Murase:2013ffa}
\bibinfo{author}{K.~Murase} and \bibinfo{author}{K.~Ioka},
  \bibinfo{journal}{Phys.Rev.Lett.} \bibinfo{volume}{\textbf{111}}(12),
  \bibinfo{pages}{121102} (\bibinfo{date}{2013}), \eprint{1306.2274}.
\bibitem{Kimura:2014jba}
\bibinfo{author}{S.~S. Kimura}, \bibinfo{author}{K.~Murase}, and
  \bibinfo{author}{K.~Toma}, \bibinfo{journal}{Astrophys.J.}
  \bibinfo{volume}{\textbf{806}}, \bibinfo{pages}{159} (\bibinfo{date}{2015}),
  \eprint{1411.3588}.
\bibitem{Senno:2015tsn}
\bibinfo{author}{N.~Senno}, \bibinfo{author}{K.~Murase}, and
  \bibinfo{author}{P.~M\'esz\'aros}, \bibinfo{journal}{Phys. Rev.}
  \bibinfo{volume}{\textbf{D93}}(8), \bibinfo{pages}{083003}
  (\bibinfo{date}{2016}), \eprint{1512.08513}.
\bibitem{2016PhRvD..93h3005W}
\bibinfo{author}{X.-Y. Wang} and \bibinfo{author}{R.-Y. Liu},
  \bibinfo{journal}{Phys. Rev. D} \bibinfo{volume}{\textbf{93}}(8),
  \bibinfo{pages}{083005} (\bibinfo{date}{Apr. 2016}).
\bibitem{2016PhRvD..93e3010T}
\bibinfo{author}{I.~Tamborra} and \bibinfo{author}{S.~Ando},
  \bibinfo{journal}{Phys. Rev. D} \bibinfo{volume}{\textbf{93}}(5),
  \bibinfo{pages}{053010} (\bibinfo{date}{Mar. 2016}).
\bibitem{Senno:2016bso}
\bibinfo{author}{N.~Senno}, \bibinfo{author}{K.~Murase}, and
  \bibinfo{author}{P.~Meszaros}, \bibinfo{journal}{Astrophys. J.}
  \bibinfo{volume}{\textbf{838}}(1), \bibinfo{pages}{3} (\bibinfo{date}{2017}),
  \eprint{1612.00918}.
\bibitem{2016arXiv161200011D}
\bibinfo{author}{L.~Dai} and \bibinfo{author}{K.~Fang},
  \bibinfo{journal}{arXiv} \bibinfo{pages}{p. arXiv:1612.00011}
  (\bibinfo{date}{Nov. 2016}), \eprint{1612.00011}.
\bibitem{Lunardini:2016wh}
\bibinfo{author}{C.~Lunardini} and \bibinfo{author}{W.~Winter},
  \bibinfo{journal}{arXiv}  (\bibinfo{date}{Dec. 2016}), \eprint{1612.03160v2}.
\bibitem{1998Natur.395..670G}
\bibinfo{author}{T.~J. Galama}, \bibinfo{author}{P.~M. Vreeswijk},
  \bibinfo{author}{J.~van Paradijs}, \bibinfo{author}{C.~Kouveliotou},
  \bibinfo{author}{T.~Augusteijn}, \bibinfo{author}{O.~R. Hainaut},
  \bibinfo{author}{F.~Patat}, \bibinfo{author}{H.~Boehnhardt},
  \bibinfo{author}{J.~Brewer}, \bibinfo{author}{V.~Doublier}, \emph{et~al.},
  \bibinfo{journal}{arXiv} (6), \bibinfo{pages}{670} (\bibinfo{date}{Jun.
  1998}), \eprint{astro-ph/9806175v1}.
\bibitem{1999ApJ...516..788W}
\bibinfo{author}{S.~E. Woosley}, \bibinfo{author}{R.~G. Eastman}, and
  \bibinfo{author}{B.~P. Schmidt}, \bibinfo{journal}{arXiv} (2),
  \bibinfo{pages}{788} (\bibinfo{date}{Jun. 1998}),
  \eprint{astro-ph/9806299v1}.
\bibitem{2011arXiv1112.5949B}
\bibinfo{author}{O.~Bromberg}, \bibinfo{author}{E.~Nakar},
  \bibinfo{author}{T.~Piran}, and \bibinfo{author}{R.~Sari},
  \bibinfo{journal}{arXiv} \bibinfo{pages}{p. 5949} (\bibinfo{date}{Dec.
  2011}), \eprint{1112.5949}.
\bibitem{Mizuta:2013he}
\bibinfo{author}{A.~Mizuta} and \bibinfo{author}{K.~Ioka},
  \bibinfo{journal}{ApJ} \bibinfo{volume}{\textbf{777}}(2),
  \bibinfo{pages}{162} (\bibinfo{date}{Nov. 2013}).
\bibitem{2013ApJ...778...18M}
\bibinfo{author}{R.~Margutti}, \bibinfo{author}{A.~M. Soderberg},
  \bibinfo{author}{M.~H. Wieringa}, \bibinfo{author}{P.~G. Edwards},
  \bibinfo{author}{R.~A. Chevalier}, \bibinfo{author}{B.~J. Morsony},
  \bibinfo{author}{R.~Barniol~Duran}, \bibinfo{author}{L.~Sironi},
  \bibinfo{author}{B.~A. Zauderer}, \bibinfo{author}{D.~Milisavljevic},
  \emph{et~al.}, \bibinfo{journal}{ApJ} \bibinfo{volume}{\textbf{778}}(1),
  \bibinfo{pages}{18} (\bibinfo{date}{Nov. 2013}).
\bibitem{Nomoto:2001uv}
\bibinfo{author}{K.~Nomoto}, \bibinfo{author}{P.~Mazzali}, and
  \bibinfo{author}{T.~Nakamura}, \bibinfo{journal}{and gamma-ray}
  (\bibinfo{date}{Jan. 2001}).
\bibitem{2016ApJ...832..108M}
\bibinfo{author}{M.~Modjaz}, \bibinfo{author}{Y.~Q. Liu},
  \bibinfo{author}{F.~B. Bianco}, and \bibinfo{author}{O.~Graur},
  \bibinfo{journal}{ApJ} \bibinfo{volume}{\textbf{832}}(2),
  \bibinfo{pages}{108} (\bibinfo{date}{Dec. 2016}).
\bibitem{2017MNRAS.472..616S}
\bibinfo{author}{E.~{Sobacchi}}, \bibinfo{author}{J.~{Granot}},
  \bibinfo{author}{O.~{Bromberg}}, and \bibinfo{author}{M.~C. {Sormani}},
  \bibinfo{journal}{2017} .
\bibitem{2017MNRAS.472.1770B}
\bibinfo{author}{E.~{Bear}}, \bibinfo{author}{A.~{Grichener}}, and
  \bibinfo{author}{N.~{Soker}}, \bibinfo{journal}{2017} .
\bibitem{Murase:2006mm}
\bibinfo{author}{K.~Murase}, \bibinfo{author}{K.~Ioka},
  \bibinfo{author}{S.~Nagataki}, and \bibinfo{author}{T.~Nakamura},
  \bibinfo{journal}{Astrophys.J.} \bibinfo{volume}{\textbf{651}},
  \bibinfo{pages}{L5} (\bibinfo{date}{2006}), \eprint{astro-ph/0607104}.
\bibitem{Gupta:2006jm}
\bibinfo{author}{N.~Gupta} and \bibinfo{author}{B.~Zhang},
  \bibinfo{journal}{Astropart.Phys.} \bibinfo{volume}{\textbf{27}},
  \bibinfo{pages}{386} (\bibinfo{date}{2007}), \eprint{astro-ph/0606744}.
\bibitem{Kashiyama:2012zn}
\bibinfo{author}{K.~Kashiyama}, \bibinfo{author}{K.~Murase},
  \bibinfo{author}{S.~Horiuchi}, \bibinfo{author}{S.~Gao}, and
  \bibinfo{author}{P.~M\'esz\'aros}, \bibinfo{journal}{Astrophys.J.}
  \bibinfo{volume}{\textbf{769}}, \bibinfo{pages}{L6} (\bibinfo{date}{2013}),
  \eprint{1210.8147}.
\bibitem{Abbasi:2011hk}
\bibinfo{author}{R.~Abbasi}, \bibinfo{author}{Y.~Abdou},
  \bibinfo{author}{T.~Abu-Zayyad}, \bibinfo{author}{J.~Adams},
  \bibinfo{author}{J.~A. Aguilar}, \bibinfo{author}{M.~Ahlers},
  \bibinfo{author}{K.~Andeen}, \bibinfo{author}{J.~Auffenberg},
  \bibinfo{author}{X.~Bai}, \bibinfo{author}{M.~Baker}, \emph{et~al.},
  \bibinfo{journal}{Phys. Rev. Lett.} \bibinfo{volume}{\textbf{106}}(14),
  \bibinfo{pages}{141101} (\bibinfo{date}{Apr. 2011}).
\bibitem{2012Natur.484..351A}
\bibinfo{author}{R.~Abbasi}, \bibinfo{author}{Y.~Abdou},
  \bibinfo{author}{T.~Abu-Zayyad}, \bibinfo{author}{M.~Ackermann},
  \bibinfo{author}{J.~Adams}, \bibinfo{author}{J.~A. Aguilar},
  \bibinfo{author}{M.~Ahlers}, \bibinfo{author}{D.~Altmann},
  \bibinfo{author}{K.~Andeen}, \bibinfo{author}{J.~Auffenberg}, \emph{et~al.},
  \bibinfo{journal}{Nature} \bibinfo{volume}{\textbf{484}}(7),
  \bibinfo{pages}{351} (\bibinfo{date}{Apr. 2012}).
\bibitem{Aartsen:2015jp}
\bibinfo{author}{M.~G. Aartsen}, \bibinfo{author}{M.~Ackermann},
  \bibinfo{author}{J.~Adams}, \bibinfo{author}{J.~A. Aguilar},
  \bibinfo{author}{M.~Ahlers}, \bibinfo{author}{M.~Ahrens},
  \bibinfo{author}{D.~Altmann}, \bibinfo{author}{T.~Anderson},
  \bibinfo{author}{C.~Arguelles}, \bibinfo{author}{T.~C. Arlen}, \emph{et~al.},
  \bibinfo{journal}{ApJ} \bibinfo{volume}{\textbf{805}}(1), \bibinfo{pages}{L5}
  (\bibinfo{date}{May 2015}).
\bibitem{2016ApJ...824..115A}
\bibinfo{author}{M.~G. Aartsen}, \bibinfo{author}{K.~Abraham},
  \bibinfo{author}{M.~Ackermann}, \bibinfo{author}{J.~Adams},
  \bibinfo{author}{J.~A. Aguilar}, \bibinfo{author}{M.~Ahlers},
  \bibinfo{author}{M.~Ahrens}, \bibinfo{author}{D.~Altmann},
  \bibinfo{author}{T.~Anderson}, \bibinfo{author}{I.~Ansseau}, \emph{et~al.},
  \bibinfo{journal}{ApJ} \bibinfo{volume}{\textbf{824}}(2),
  \bibinfo{pages}{115} (\bibinfo{date}{Jun. 2016}).
\bibitem{2017arXiv170206868A}
\bibinfo{author}{M.~G. Aartsen}, \bibinfo{author}{M.~Ackermann},
  \bibinfo{author}{J.~Adams}, \bibinfo{author}{J.~A. Aguilar},
  \bibinfo{author}{M.~Ahlers}, \bibinfo{author}{M.~Ahrens},
  \bibinfo{author}{I.~Al~Samarai}, \bibinfo{author}{D.~Altmann},
  \bibinfo{author}{K.~Andeen}, \bibinfo{author}{T.~Anderson}, \emph{et~al.},
  \bibinfo{journal}{arXiv} \bibinfo{pages}{p. arXiv:1702.06868}
  (\bibinfo{date}{Feb. 2017}), \eprint{1702.06868}.
\bibitem{2001PhRvL..87q1102M}
\bibinfo{author}{P.~M{\'e}sz{\'a}ros} and \bibinfo{author}{E.~Waxman},
  \bibinfo{journal}{Phys. Rev. Lett.} \bibinfo{volume}{\textbf{87}}(1),
  \bibinfo{pages}{171102} (\bibinfo{date}{Oct. 2001}).
\bibitem{Razzaque:2003eq}
\bibinfo{author}{S.~Razzaque}, \bibinfo{author}{P.~M{\'e}sz{\'a}ros}, and
  \bibinfo{author}{E.~Waxman}, \bibinfo{journal}{Phys. Rev. D}
  \bibinfo{volume}{\textbf{68}}(8), \bibinfo{pages}{083001}
  (\bibinfo{date}{Oct. 2003}).
\bibitem{Razzaque:2004ix}
\bibinfo{author}{S.~Razzaque}, \bibinfo{author}{P.~M{\'e}sz{\'a}ros}, and
  \bibinfo{author}{E.~Waxman}, \bibinfo{journal}{Phys. Rev. Lett.}
  \bibinfo{volume}{\textbf{93}}(18), \bibinfo{pages}{181101}
  (\bibinfo{date}{Oct. 2004}).
\bibitem{2005MPLA...20.2351R}
\bibinfo{author}{S.~Razzaque}, \bibinfo{author}{P.~M{\'e}sz{\'a}ros}, and
  \bibinfo{author}{E.~Waxman}, \bibinfo{journal}{Modern Physics Letters A}
  \bibinfo{volume}{\textbf{20}}(3), \bibinfo{pages}{2351} (\bibinfo{date}{Jan.
  2005}).
\bibitem{Ando:2005xi}
\bibinfo{author}{S.~Ando} and \bibinfo{author}{J.~F. Beacom},
  \bibinfo{journal}{Phys. Rev. Lett.} \bibinfo{volume}{\textbf{95}},
  \bibinfo{pages}{061103} (\bibinfo{date}{2005}), \eprint{astro-ph/0502521}.
\bibitem{Iocco:2007td}
\bibinfo{author}{F.~Iocco}, \bibinfo{author}{K.~Murase},
  \bibinfo{author}{S.~Nagataki}, and \bibinfo{author}{P.~D. Serpico},
  \bibinfo{journal}{Astrophys. J.} \bibinfo{volume}{\textbf{675}},
  \bibinfo{pages}{937} (\bibinfo{date}{2008}), \eprint{0707.0515}.
\bibitem{Nakar:2015tma}
\bibinfo{author}{E.~Nakar}, \bibinfo{journal}{Astrophys. J.}
  \bibinfo{volume}{\textbf{807}}(2), \bibinfo{pages}{172}
  (\bibinfo{date}{2015}), \eprint{1503.00441}.
\bibitem{2011A&A...527A..28I}
\bibinfo{author}{{IceCube Collaboration}}, \bibinfo{author}{R.~{Abbasi}},
  \bibinfo{author}{Y.~{Abdou}}, \bibinfo{author}{T.~{Abu-Zayyad}},
  \bibinfo{author}{J.~{Adams}}, \bibinfo{author}{J.~A. {Aguilar}},
  \bibinfo{author}{M.~{Ahlers}}, \bibinfo{author}{K.~{Andeen}},
  \bibinfo{author}{J.~{Auffenberg}}, \bibinfo{author}{X.~{Bai}}, \emph{et~al.},
  \bibinfo{journal}{Astronomy and Astrophysics} \bibinfo{volume}{\textbf{527}},
  \bibinfo{pages}{A28}, \bibinfo{eid}{A28} (\bibinfo{date}{Mar. 2011}),
  \eprint{1101.3942}.
\bibitem{2009PASP..121.1334R}
\bibinfo{author}{A.~Rau}, \bibinfo{author}{S.~R. Kulkarni},
  \bibinfo{author}{N.~M. Law}, \bibinfo{author}{J.~S. Bloom},
  \bibinfo{author}{D.~Ciardi}, \bibinfo{author}{G.~S. Djorgovski},
  \bibinfo{author}{D.~B. Fox}, \bibinfo{author}{A.~Gal-Yam},
  \bibinfo{author}{C.~C. Grillmair}, \bibinfo{author}{M.~M. Kasliwal},
  \emph{et~al.}, \bibinfo{journal}{Publications of the Astronomical Society of
  Pacific} \bibinfo{volume}{\textbf{121}}(8), \bibinfo{pages}{1334}
  (\bibinfo{date}{Dec. 2009}).
\bibitem{Smith:2012eu}
\bibinfo{author}{M.~W.~E. Smith} \emph{et~al.}, \bibinfo{journal}{Astropart.
  Phys.} \bibinfo{volume}{\textbf{45}}, \bibinfo{pages}{56}
  (\bibinfo{date}{2013}), \eprint{1211.5602}.
\bibitem{2010RScI...81h1101H}
\bibinfo{author}{F.~Halzen} and \bibinfo{author}{S.~R. Klein},
  \bibinfo{journal}{Review of Scientific Instruments}
  \bibinfo{volume}{\textbf{81}}(8), \bibinfo{pages}{081101}
  (\bibinfo{date}{Aug. 2010}).
\bibitem{Guillochon:2016ci}
\bibinfo{author}{J.~Guillochon}, \bibinfo{author}{J.~Parrent},
  \bibinfo{author}{L.~Z. Kelley}, and \bibinfo{author}{R.~Margutti},
  \bibinfo{journal}{arXiv}  (\bibinfo{date}{May 2016}), \eprint{1605.01054v2}.
\bibitem{2008APh....29..299B}
\bibinfo{author}{J.~Braun}, \bibinfo{author}{J.~Dumm},
  \bibinfo{author}{F.~De~Palma}, \bibinfo{author}{C.~Finley},
  \bibinfo{author}{A.~Karle}, and \bibinfo{author}{T.~Montaruli},
  \bibinfo{journal}{Astroparticle Physics} \bibinfo{volume}{\textbf{29}}(4),
  \bibinfo{pages}{299} (\bibinfo{date}{May 2008}).
\bibitem{2017AdAst2017E...5C}
\bibinfo{author}{Z.~Cano}, \bibinfo{author}{S.-Q. Wang}, \bibinfo{author}{Z.-G.
  Dai}, and \bibinfo{author}{X.-F. Wu}, \bibinfo{journal}{Advances in
  Astronomy} \bibinfo{volume}{\textbf{2017}}(1), \bibinfo{pages}{8929054}
  (\bibinfo{date}{2017}).
\bibitem{kent1982fisher}
\bibinfo{author}{J.~T. Kent}, \bibinfo{journal}{Journal of the Royal
  Statistical Society. Series B (Methodological)} \bibinfo{pages}{pp. 71--80}
  (\bibinfo{date}{1982}).
\bibitem{2016ApJ...816...75A}
\bibinfo{author}{M.~G. Aartsen}, \bibinfo{author}{K.~Abraham},
  \bibinfo{author}{M.~Ackermann}, \bibinfo{author}{J.~Adams},
  \bibinfo{author}{J.~A. Aguilar}, \bibinfo{author}{M.~Ahlers},
  \bibinfo{author}{M.~Ahrens}, \bibinfo{author}{D.~Altmann},
  \bibinfo{author}{T.~Anderson}, \bibinfo{author}{I.~Ansseau}, \emph{et~al.},
  \bibinfo{journal}{ApJ} \bibinfo{volume}{\textbf{816}}(2), \bibinfo{pages}{75}
  (\bibinfo{date}{Jan. 2016}).
\bibitem{1999PhRvD..59b3002W}
\bibinfo{author}{E.~Waxman} and \bibinfo{author}{J.~Bahcall},
  \bibinfo{journal}{Physical Review D (Particles}
  \bibinfo{volume}{\textbf{59}}(2), \bibinfo{pages}{023002}
  (\bibinfo{date}{Jan. 1999}).
\bibitem{Aartsen:2014cva}
\bibinfo{author}{M.~G. Aartsen} \emph{et~al.} (\bibinfo{collaboration}{IceCube
  Collaboration}), \bibinfo{journal}{Astrophys. J.}
  \bibinfo{volume}{\textbf{796}}(2), \bibinfo{pages}{109}
  (\bibinfo{date}{2014}), \eprint{1406.6757}.
\bibitem{1986ApJ...303..336G}
\bibinfo{author}{N.~Gehrels}, \bibinfo{journal}{Astrophysical Journal}
  \bibinfo{volume}{\textbf{303}}, \bibinfo{pages}{336} (\bibinfo{date}{Apr.
  1986}).
\bibitem{Murase:2013hh}
\bibinfo{author}{K.~Murase}, \bibinfo{author}{K.~Kashiyama}, and
  \bibinfo{author}{P.~Mészáros}, \bibinfo{journal}{Phys. Rev. Lett.}
  \bibinfo{volume}{\textbf{111}}, \bibinfo{pages}{131102}
  (\bibinfo{date}{2013}), \eprint{1301.4236}.
\bibitem{Kashiyama:2013ata}
\bibinfo{author}{K.~Kashiyama}, \bibinfo{author}{K.~Murase}, and
  \bibinfo{author}{P.~Mészáros}, \bibinfo{journal}{Phys. Rev. Lett.}
  \bibinfo{volume}{\textbf{111}}, \bibinfo{pages}{131103}
  (\bibinfo{date}{2013}), \eprint{1304.1945}.
\bibitem{Bellm:2014tu}
\bibinfo{author}{E.~C. Bellm}, \bibinfo{journal}{arXiv}  (\bibinfo{date}{Oct.
  2014}), \eprint{1410.8185v1}.

\end{thebibliography}

\end{document}